\documentclass[twocolumn,epjc3]{svjour3}

\RequirePackage[T1]{fontenc}

\smartqed  

\RequirePackage{graphicx}
\RequirePackage{mathptmx}      
\RequirePackage{flushend}
\RequirePackage[numbers,sort&compress]{natbib}
\RequirePackage[colorlinks,citecolor=blue,urlcolor=blue,linkcolor=blue]{hyperref}

\usepackage{amsmath}
\usepackage{graphicx}
\usepackage{amssymb}
\usepackage{bm}
\usepackage{array}
\usepackage{hyperref}
\usepackage{slashed}
\usepackage{braket}

\journalname{Eur. Phys. J. C}

\begin{document}

\title{Azimuthal asymmetries in lepton-pair production at a fixed-target experiment using the LHC beams (AFTER)
}


\author{Tianbo Liu\thanksref{addr1}
        \and
        Bo-Qiang Ma\thanksref{e1,addr1,addr2} 
}

\thankstext{e1}{e-mail: mabq@pku.edu.cn}

\institute{School of Physics and State Key Laboratory of Nuclear
Physics and Technology, Peking University, Beijing 100871, China\label{addr1}
          \and
          Center for High Energy Physics, Peking University, Beijing
100871, China\label{addr2}
}

\date{Received: date / Accepted: date}

\maketitle

\begin{abstract}
A multi-purpose fixed-target experiment using the proton and
lead-ion beams of the LHC was recently proposed by Brodsky, Fleuret,
Hadjidakis and Lansberg, and here we concentrate our study on some
issues related to the spin physics part of this project (referred to
as AFTER). We study the nucleon spin structure through $pp$ and $pd$
processes with a fixed-target experiment using the LHC proton beams,
for the kinematical region with 7~TeV proton beams at the energy in
center-of-mass frame of two nucleons $\sqrt{s}=115$ GeV. We
calculate and estimate the $\cos2\phi$ azimuthal asymmetries of
unpolarized $pp$ and $pd$ dilepton production processes in the
Drell--Yan continuum region and at
the $Z$-pole. We also calculate the $\sin(2\phi-\phi_S)$,
$\sin(2\phi+\phi_S)$ and $\sin2\phi$ azimuthal asymmetries of $pp$
and $pd$ dilepton production processes with the target proton and
deuteron longitudinally or transversally polarized in the Drell--Yan
continuum region and around $Z$
resonances region. We conclude that it is feasible to measure these
azimuthal asymmetries, consequently the three-dimensional or
transverse momentum dependent parton distribution functions (3dPDFs
or TMDs), at this new AFTER facility.
\end{abstract}

\section{Introduction}

Recently a multi-purpose fixed-target experiment using the proton
and lead-ion beams of the Large Hadron Collider (LHC) extracted by a
bent crystal, referred to as AFTER in the following, was proposed by
Brodsky, Fleuret, Hadjidakis and Lansberg~\cite{Brodsky:2012vg}.
Such an extraction mode will not alter the performance of the
collider experiments at the LHC. The center-of-mass energy is
$\sqrt{s_{NN}}=115$ GeV with the LHC-$7$ TeV proton beam and
$\sqrt{s_{NN}}=72$ GeV with the a lead running with $2.76$
TeV-per-nucleon beam, and it can be even higher by using the Fermi
motion of the nucleons in a nuclear target. This project will
provide a unique opportunity to study the nucleon partonic
structure, spin physics, nuclear matter properties, deconfinement in
heavy ion collisions, $W$ and $Z$ productions, exclusive,
semi-exclusive and backward reactions, and even further
potentialities of a high-energy fixed target set-up. We concentrate
our study on some issues related to the spin physics for the AFTER
proposal.

The study of the three-dimensional or the intrinsic transverse
momentum dependent distribution functions (3dPDFs or TMDs) has
received much attention in recent years~\cite{Barone:2001sp}. Such
new quantities of the nucleon provide us a significant perspective
on understanding the three-dimensional structure of hadrons and the
non-perturbative properties of quantum chromodynamics (QCD). The
intrinsic transversal momentum of partons may cause special effects
in high energy scattering experiments~\cite{Cahn:1978se}. Azimuthal
asymmetries of unpolarized and single polarized Drell--Yan processes
are among the most challenging issues of
QCD spin physics~\cite{D'Alesio:2007jt,Barone:2010zz,Boer:2011fh}.

The first measurement of the angular distribution of Drell--Yan process,
performed by NA10 Collaboration for $\pi N$, indicates a sizable $\cos2\phi$
azimuthal asymmetry~\cite{Falciano:1986wk,Guanziroli:1987rp} which cannot be
described by leading and next-to-leading order perturbative QCD~\cite{Brandenburg:1993cj}.
Furthermore, the violation of the Lam--Tung relation~\cite{Lam:1978pu} which is obtained from
the spin-$1/2$ nature of quarks and the spin-$1$ nature of gluons,
just like the Callan--Gross relation in the deep-inelastic scattering~\cite{Callan:1969uq},
was measured by Fermilab E615 Collaboration~\cite{Conway:1989fs}.
This violation was also tested by E866/NuSea Collaboration through
the $pd$ and $pp$ Drell--Yan dimuon processes in recent year~\cite{Zhu:2006gx,Zhu:2008sj}.

Large single spin asymmetries (SSAs) were observed experimentally in the process $pp^\uparrow\rightarrow\pi X$
two decades ago~\cite{Adams:1991rw,Adams:1991rv,Adams:1991ru,Adams:1991cs,Adams:1994yu,Bravar:1996ki}.
SSAs in semi-inclusive deeply inelastic scattering (SIDIS)~\cite{Bravar:2000ti,Airapetian:2004tw,Alexakhin:2005iw,Diefenthaler:2005gx,Airapetian:2009ae,Ageev:2006da,Alekseev:2010rw}
with one colliding nucleon transversely polarized have also been measured by several experiments.
Standard perturbtive QCD based on collinear factorization to leading twist failed to explain these
asymmetries~\cite{Kane:1978nd}.

The Drell--Yan process is an ideal ground for testing the
perturbative QCD and for probing the 3dPDFs or TMDs, as it contains
only the distribution functions with no fragmentation functions, and
its differential cross section is well described by next-to-leading
order QCD calculations~\cite{Stirling:1993gc}. In this paper, we
calculate azimuthal asymmetries of $pp$ and $pd$ dilepton production
processes in Drell--Yan continuum region and around the $Z$-pole through a fixed-target experiment
using the LHC proton beams with the proton or deuteron target
unpolarized and transversally or longitudinally polarized. The paper
is organized as follows. In Sect.~2 and 3, we respectively calculate
the azimuthal asymmetries in unpolarized and single polarized $pp$
and $pd$ processes. In Sect.~4, we present the numerical results of
these asymmetries. Then, a brief discussion and conclusion is
contained in Sect.~5.

\section{The $\cos2\phi$ azimuthal asymmetries of unpolarized $pp$ and $pd$ processes}

The Drell--Yan process is an ideal ground to investigate the hadron structure,
because it only probes the parton distributions without fragmentation functions.
It was naively speculated that the polarization of at least one incoming
hadron is necessary to investigate the spin-related structure and properties of hadrons.
However, it is not the case if we take the intrinsic transversal momentum of quarks
inside the hadron into account.
As mentioned before, the standard perturbative QCD to leading and next-to-leading order failed to
describe the sizable $\cos2\phi$ azimuthal asymmetry and the Lam--Tung relation violation of the
unpolarized Drell--Yan experiments~\cite{Falciano:1986wk,Guanziroli:1987rp,Conway:1989fs,Zhu:2006gx,Zhu:2008sj}.
Several attempts were made to interpret this asymmetry, such as the factorization breaking
QCD vacuum effect~\cite{Brandenburg:1993cj} (which corresponds possibly the helicity flip in the instanton
model~\cite{Boer:2004mv}), higher twist effect~\cite{Brandenburg:1994wf,Eskola:1994py,Heinrich:1991zm}
and the coherent states~\cite{Blazek:1989kt}. Boer pointed out that the $\cos2\phi$ azimuthal
asymmetry could be due to a non-vanished 3dPDF or TMD $h_1^\perp(x,\bm{p}_T^2)$~\cite{Boer:1999mm},
named as the Boer--Mulders function later, as one of the eight leading-twist 3dPDFs or TMD distribution
functions contained in~\cite{Mulders:1995dh,Boer:1997nt}
\begin{equation}\label{tmds}
\begin{split}
\Phi = & \frac{1}{2} \bigg\{ f_1\slashed{n}_+ - f_{1T}^\perp \frac{\epsilon_T^{ij} p_{Ti} S_{Tj}}{M}\slashed{n}_+  + h_{1T}\frac{[\slashed{S}_T,\slashed{n}_+]\gamma_5}{2} \\
& + \Big(S_L g_{1L} + \frac{\bm{p}_T \cdot \bm{S}_T}{M} g_{1T}\Big) \gamma_5\slashed{n}_+ \\
&  + \Big(S_L h_{1L}^\perp + \frac{\bm{p}_T \cdot \bm{S}_T}{M} h_{1T}^\perp\Big) \frac{[\slashed{p}_T,\slashed{n}_+]\gamma_5}{2M} \\
& + i h_1^\perp \frac{[\slashed{p}_T,\slashed{n}_+]}{2M} \bigg\},
\end{split}
\end{equation}
where $\Phi$ is the quark-quark correlation matrix, defined as
\begin{equation}
\Phi_{ij}(p, P, S) = \int \frac{d^4\xi}{(2\pi)^4} e^{ip\cdot \xi} \braket{PS | \bar{\psi}_j(0) \mathcal{W}[0,\xi]\psi_i(\xi) | PS}.
\end{equation}
The sizable $\cos2\phi$ azimuthal asymmetry can arise from a product of two Boer--Mulders functions
of two incoming hadrons by establishing a preferred transverse momentum direction from the spin--transverse
momentum correlation. This effect is called the Boer--Mulders effect~\cite{Boer:1999mm}.
Many theoretical and phenomenological studies
are carried out along this
direction~\cite{Lu:2004hu,Lu:2005rq,Bianconi:2005bd,Sissakian:2005vd,Sissakian:2005yp,Lu:2006ew,Barone:2006ws,Lu:2007kj,Gamberg:2005ip,Zhang:2008nu,Zhang:2008ez,Barone:2010gk,Lu:2009ip,Lu:2011mz,Yuan:2003wk,Pasquini:2006iv,Gockeler:2006zu,Burkardt:2007xm,Liu:2012fh}.

The Boer--Mulders function $h_1^\perp$, as well as the Sivers function $f_{1T}^\perp$,
is a naively time-reversal odd ($T$-odd) distribution function, characterizing the correlation
between quark transversal momentum and quark transversal spin.
Therefore, it was thought to be forbidden for a long time because of the time-reversal invariance
property of QCD~\cite{Collins:1992kk}. However the model calculations taken by Brodsky, Hwang and
Schmidt indicated that these non-vanished naively $T$-odd distribution functions, $h_1^\perp$ and $f_{1T}^\perp$ can arise from the final or initial state interaction between the struck quark and the target remnant in the SIDIS and Drell--Yan processes at leading-twist level~\cite{Brodsky:2002cx,Brodsky:2002rv}.
In general, the path-order Wilson line, which arises from the requirement of a full QCD gauge invarient definition
of 3dPDFs or TMD distribution functions, provides non-trivial phases and leads to non-vanished $T$-odd distribution functions~\cite{Ellis:1982wd,Efremov:1979qk,Collins:1981uw,Ji:2002aa,Belitsky:2002sm,Boer:2003cm}.
Due to the present of the Wilson line, opposite sign of Boer--Mulders function or Sivers function
in SIDIS and Drell--Yan processes is expected~\cite{Boer:2003cm,Collins:2002kn,Collins:2004nx}
\begin{eqnarray}
h_1^\perp(x, \bm{p}_T^2)|_{\mathrm{SIDIS}} = - h_1^\perp(x, \bm{p}_T^2)|_{\mathrm{DY}},\\
f_{1T}^\perp(x, \bm{p}_T^2)|_{\mathrm{SIDIS}} = - f_{1T}^\perp(x, \bm{p}_T^2)|_{\mathrm{DY}}.
\end{eqnarray}
This relation still awaits for experimental confirmation. For hadron productions in hadron--hadron
collisions, the situation is more involved, since colored objects exist in both the initial
and the final states. The multiple final or initial state interactions will generate
process-dependent 3dPDFs or TMDs which may be different from those in SIDIS or
Drell--Yan process~\cite{Collins:2007nk,Collins:2007jp,Vogelsang:2007jk,Bomhof:2007xt}.
This is also viewed as the breakdown of the generalized 3dPDF or TMD factorization in the inclusive
hadro-production of hadrons~\cite{Rogers:2010dm}.

\begin{figure}
  \includegraphics[width=0.45\textwidth]{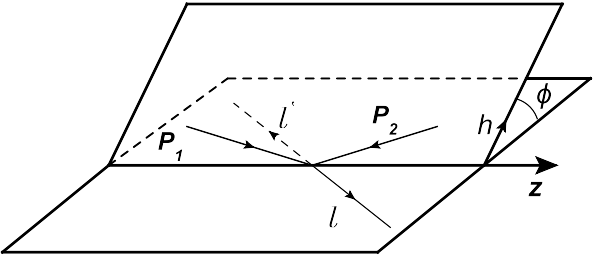}
\caption{The Collins--Soper frame.}
\label{csframe}       
\end{figure}

The angular differential cross section for unpolarized Drell--Yan process has
the general form:
\begin{align}\label{uncrosssection}
\frac{1}{\sigma}\frac{d\sigma}{d\Omega}=&\frac{3}{4\pi}\frac{1}{\lambda+3}\bigg(1+\lambda\cos^2\theta
+\mu\sin2\theta\cos\phi\nonumber\\
&+\frac{\nu}{2}\sin^2\theta\cos2\phi\bigg),
\end{align}
where $\Omega$ is the solid angle and $\lambda$, $\mu$, and $\nu$
are angular distribution coefficients. For azimuthal symmetrical
scattering, the coefficients $\mu=\nu=0$. The polar and azimuthal
angular $\theta$ and $\phi$ are defined in the Collins--Soper (CS)
frame~\cite{Collins:1977iv}, as shown in Fig.~\ref{csframe}. It is
the center of mass of the lepton pair with the $z$ axis defined as
the bisector of two incoming hadrons. The polar angular $\theta$ is
defined as the angular of the positive lepton with respect to the
$z$ axis direction, and the azimuthal angular $\phi$ is defined as
the angular of the lepton plane with respect to the proton plane. In
this frame the Lam--Tung relation is insensitive to the higher
fixed-order perturbative QCD~\cite{Mirkes:1994eb} or the QCD
resummation~\cite{Boer:2006eq,Berger:2007si,Berger:2007jw}.

Taking into account the Boer--Mulders distribution, we can express the unpolarized Drell--Yan
cross section as
\begin{equation}
\begin{split}\label{undrellyan}
\frac{d\sigma}{d\Omega dx_1dx_2d^2\bm{q}_T}=\frac{\alpha_{\textrm{em}}^2}{12Q^2}\sum_a e_a^2
\bigg\{(1+\cos^2\theta)\mathcal{F}[f_{1a}\bar{f}_{1a}]\\
+\sin^2\theta\cos2\phi\mathcal{F}\bigg[\frac{2\hat{\bm{h}}\cdot\bm{p}_T\hat{\bm{h}}\cdot\bm{k}_T-\bm{p}_T\cdot\bm{k}_T}{m_N^2} h_{1a}^\perp\bar{h}_{1a}^\perp\bigg]\bigg\},
\end{split}
\end{equation}
where $\hat{\bm{h}}\equiv\bm{q}_T/|\bm{q}_T|$ is the direction of the transversal momentum transfer,
$\bm{p}_T$ and $\bm{k}_T$ are the transversal momentum of quarks in the nucleons, $m_N$ is the mass of a nucleon,
$\alpha_{em}$ is the electromagnetic fine structure constant and $e_a$ is the charge of the quark
with the subscript $a$ showing the flavor.
The structure function notation in the equation is defined as
\begin{equation}\label{Fnotation}
\mathcal{F}[\cdots]=\int d^2\bm{p}_T d^2\bm{k}_T\delta^2(\bm{p}_T+\bm{k}_T-\bm{q}_T)[\cdots].
\end{equation}
Then the $\cos2\phi$ azimuthal asymmetry can be expressed as
\begin{equation}\label{nu1}
\nu=\frac{2\frac{1}{Q^2}\sum_a e_a^2\mathcal{F}\left[\frac{2\hat{\bm{h}}\cdot\bm{p}_T\hat{\bm{h}}\cdot\bm{k}_T-\bm{p}_T\cdot\bm{k}_T}{m_N^2} h_{1a}^\perp\bar{h}_{1a}^\perp\right]}{\frac{1}{Q^2}\sum_a e_a^2\mathcal{F}[f_{1a}\bar{f}_{1a}]}.
\end{equation}

If we take both $\gamma^*$ and $Z$ boson into account, the cross section is expressed as~\cite{Boer:1999mm}
\begin{equation}\label{diffBoer}
\begin{split}
\frac{d\sigma}{d\Omega dx_1dx_2d^2\bm{q}_T}&=\frac{\alpha_{\textrm{em}}^2}{3Q^2}
\sum_a\bigg\{K_1(\theta)\mathcal{F}[f_{1a}\bar{f}_{1a}]\\
&+[K_3(\theta)\cos2\phi+K_4(\theta)\sin2\phi]\\
&\times\mathcal{F}
\bigg[(2\hat{\bm{h}}\cdot\bm{p}_T\hat{\bm{h}}\cdot\bm{k}_T-\bm{p}_T\cdot\bm{k}_T)
\frac{h_{1a}^{\perp}\bar{h}_{1a}^{\perp}}{m_N^2}\bigg]\bigg\}.
\end{split}
\end{equation}
The coefficients $K_1$, $K_2$, $K_3$, $K_4$ are expressed as
\begin{eqnarray}
K_1(\theta)&=&\frac{1}{4}(1+\cos^2\theta)[e_a^2+2g^V_\ell e_ag^V_a\chi_1+c_1^\ell c_1^a\chi_2]\nonumber\\
&&+\frac{\cos\theta}{2}[2g^A_\ell e_ag^A_a\chi_1+c_3^\ell c_3^a\chi_2],\label{K1}\\
K_2(\theta)&=&\frac{1}{4}(1+\cos^2\theta)[2g^V_\ell e_ag^A_a\chi_1+c_1^\ell c_3^a\chi_2]\nonumber\\
&&+\frac{\cos\theta}{2}[2g^A_\ell e_ag^V_a\chi_1+c_3^\ell c_1^a\chi_2],\\
K_3(\theta)&=&\frac{1}{4}\sin^2\theta[e_a^2+2g^V_\ell e_ag^V_a\chi_1+c_1^\ell c_2^a\chi_2],\\
K_4(\theta)&=&\frac{1}{4}\sin^2\theta[2g^V_\ell e_ag^A_a\chi_3],\label{K4}
\end{eqnarray}
where the combinations of the coupling constants are
\begin{equation}
\begin{split}
c_1^j=({g^V_j}^2+{g^A_j}^2)&,\quad c_2^j=({g^V_j}^2-{g^A_j}^2),\\
c_3^j=2g^V_jg^A_j&,
\end{split}
\end{equation}
and the $Z$ boson propagator factors are expressed as
\begin{eqnarray}
\chi_1&=&\frac{1}{\sin^2(2\theta_W)}\frac{Q^2(Q^2-m_Z^2)}{(Q^2-m_Z^2)^2+\Gamma_Z^2m_Z^2},\\
\chi_2&=&\frac{1}{\sin^2(2\theta_W)}\frac{Q^2}{Q^2-m_Z^2}\chi_1,\\
\chi_3&=&-\frac{\Gamma_Z m_Z}{Q^2-m_Z^2}\chi_1,\label{chi3}
\end{eqnarray}
where $\theta_W$ is the Weinberg angle. Then the $\cos2\phi$ azimuthal asymmetry is
\begin{equation}
\begin{split}\label{nu3}
\nu &=\\
&\frac{2\sum_a\frac{e_a^2+2g^V_\ell e_ag^V_a\chi_1+c_1^\ell c_1^a\chi_2}{Q^2}\mathcal{F}\left[\frac{2\hat{\bm{h}}\cdot\bm{p}_T\hat{\bm{h}}\cdot\bm{k}_T-\bm{p}_T\cdot\bm{k}_T}{m_N^2} h_{1a}^\perp\bar{h}_{1a}^\perp\right]}{\sum_a\frac{e_a^2+2g^V_\ell e_ag^V_a\chi_1+c_1^\ell c_1^a\chi_2}{Q^2}\mathcal{F}[f_{1a}\bar{f}_{1a}]}.
\end{split}
\end{equation}
Another azimuthal dependent term is the $\sin2\phi$ term in Eq. (\ref{diffBoer}).
However the $\sin2\phi$ term is $1/Q^2$ suppressed. This suppression can be found from
(\ref{K4}) and (\ref{chi3}).

For $pd$ dilepton production processes, we assume the isospin
relation. The distribution functions of $u$ or $\bar{u}$ quark in
proton is the same as those of $d$ or $\bar{d}$ quark in neutron,
and the distribution functions of $d$ or $\bar{d}$ quark in proton
is the same as those of $u$ or $\bar{u}$ quark in neutron. We can
also neglect the nuclear effect of deuteron, since it is a weakly
bound state of a proton and a neutron. Therefore, for $pd$
processes, we need to replace the 3dPDFs or TMDs of the target
proton in Eq. (\ref{nu1})(\ref{nu3}) as
\begin{equation}\label{replacingdeutron}
f_u\rightarrow\frac{1}{2}(f_u+f_d),
\end{equation}
with $f$ representing $f_1$ or $h_1^\perp$ and similar for $d$, $\bar{u}$ and $\bar{d}$ quarks.
Then we can get the $\cos2\phi$ azimuthal asymmetry coefficient $\nu$ for $pd$ dilepton production
in Drell--Yan continuum region and around the $Z$ pole.

\section{The $\sin(2\phi-\phi_S)$, $\sin(2\phi+\phi_S)$ and $\sin2\phi$ azimuthal asymmetries of single polarized $pp$ and $pd$ processes}

Large SSAs observed experimentally~\cite{Adams:1991rw,Adams:1991rv,Adams:1991ru,Adams:1991cs,Adams:1994yu,Bravar:1996ki,Bravar:2000ti,Airapetian:2004tw,Alexakhin:2005iw,Diefenthaler:2005gx,Ageev:2006da,Airapetian:2009ae,Alekseev:2010rw}
cannot be interpreted by the standard perturbative QCD based on collinear factorization to leading twist.
As a challenging issue in hadron structure and QCD spin physics, many theoretical studies were proposed to explain origin of such asymmetries~\cite{Boros:1993ps,Sivers:1989cc,Sivers:1990fh,Anselmino:1994tv,Anselmino:1998yz,Yuan:2008vn}.
In the 3dPDF or TMD framework, the non-vanished naively $T$-odd Sivers function $f_{1T}^\perp$ in Eq.(\ref{tmds}),
which characterizes the correlation between quark transversal momentum and hadron transversal spin,
was applied to explain the
SSAs observed in the process $pp^\uparrow\rightarrow\pi X$~\cite{Sivers:1989cc,Sivers:1990fh,Anselmino:1994tv,Anselmino:1998yz}. SSAs contributed by this Sivers effect in SIDIS processes with one nucleon transversally polarized
have been measured by several experiments in recent years~\cite{Bravar:2000ti,Airapetian:2004tw,Alexakhin:2005iw,Ageev:2006da,Alekseev:2010rw,Airapetian:2010ds,:2009ti,Qian:2011py}.
The data on the Sivers SSAs have been utilized by different groups to extract the Sivers function
of the proton on the basis of the 3dPDF or TMD factorization~\cite{Collins:2004nx,Anselmino:2005ea,Efremov:2004tp,Collins:2005ie,Vogelsang:2005cs,Anselmino:2008sga,Ji:2004wu,Ji:2004xq}.

For a fixed-target experiment, it is convenient to polarized the target to allow the SSAs measurements.
Five leading twist 3dPDFs or TMDs in Eq. (\ref{tmds}), $f_{1T}^\perp$, $h_1^\perp$, $h_{1T}^\perp$, $g_{1T}$ and $h_{1L}^\perp$,
vanish upon integrating over the transversal momentum $\bm{k}_T$. The two naively $T$-odd distribution function
$f_{1T}^\perp$, the Sivers function, and $h_1^\perp$, the Boer--Mulders function are account for the SSAs in various processes.
Four leading twist 3dPDFs or TMDs, $h_{1T}$, $h_1^\perp$, $h_{1T}^\perp$ and $h_{1L}^\perp$, are chirally odd,
so they describe densities of the probed quarks with helicity flipped.
The $h_{1T}$ and $h_{1T}^\perp$ have the relation with the $h_1$ that
\begin{equation}
h_1(x,\bm{k}_T^2)=h_{1T}(x,\bm{k}_T^2)+\frac{\bm{k}_T^2}{2m_N^2}h_{1T}^\perp(x,\bm{k}_T^2).
\end{equation}
The distribution functions $h_1$ and $h_1^\perp$ respectively characterize the densities of transversely polarized quarks
inside a transversely polarized proton and an unpolarized proton. The distribution functions $h_{1T}^\perp$ and $h_{1L}^\perp$, arising from the double spin correlations in the parton distribution functions, respectively describe the densities of transversely polarized quarks in a transversely orthogonally polarized proton and longitudinally polarized proton.

The chiral-odd 3dPDFs or TMDs are rather difficult to be probed in high energy scattering experiments,
because they only manifest their effects by combining with another chiral-odd function, Collins fragmentation function in SIDIS or another chiral-odd distribution function in Drell--Yan.
Some efforts have been made to extract the transversity from SIDIS data~\cite{Anselmino:2007fs,Anselmino:2008jk}
and to extract the Boer--Mulders function from SIDIS and Drell--Yan data~\cite{Zhang:2008nu,Barone:2010gk,Lu:2009ip,Barone:2009hw}.
There are some extensive model calculations of $h_{1T}^\perp$ and $h_{1L}^\perp$~\cite{Pasquini:2008ax,Bacchetta:2008af,Avakian:2008dz,She:2009jq,Efremov:2009ze,Boffi:2009sh,Avakian:2010br,Zhu:2011zza,Zhu:2011ir,Zhu:2011ym,Lu:2011cw,Yuan:2008vn}.

\begin{figure}
  \includegraphics[bb=117 581 267 678]{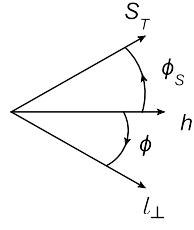}
\caption{The definition of azimuthal angles.}
\label{angle2}       
\end{figure}

If the transverse momentum of the dilepton in the Drell--Yan process $\bm{q}_T$ is measured,
we can apply the 3dPDF or TMD factorization~\cite{Collins:1981uw,Collins:2004nx,Ji:2004wu,Ji:2004xq}, which is valid in $\bm{q}_T^2\ll Q^2$ region. Then the leading order of the differential cross section can be expressed as~\cite{Boer:1999mm,Arnold:2008kf}
\begin{equation}
\begin{split}\label{spdrellyan}
&\frac{d\sigma}{d\Omega dx_1dx_2d^2\bm{q}_T}=\frac{\alpha_{\textrm{em}}}{3Q^2}\sum_ae_a^2\bigg\{\frac{1}{4}(1+\cos^2\theta)\mathcal{F}[f_{1a}\bar{f}_{1a}]\\
&+S_L\frac{\sin^2\theta}{4}\sin2\phi\mathcal{F}
\bigg[\frac{2(\hat{\bm{h}}\cdot\bm{p}_T)(\hat{\bm{h}}\cdot\bm{k}_T)-\bm{p}_T\cdot\bm{k}_T}{m_N^2}h_{1La}^\perp\bar{h}_{1a}^\perp\bigg]\\
&+|\bm{S}_T|\frac{\sin^2\theta}{4}\bigg[\sin(2\phi+\phi_S)\mathcal{F}\bigg[\big(2(\hat{\bm{h}}\cdot\bm{p}_T)(2(\hat{\bm{h}}\cdot\bm{p}_T)(\hat{\bm{h}}\cdot\bm{k}_T)\\
&\quad-\bm{p}_T\cdot\bm{k}_T)-\bm{p}_T^2(\hat{\bm{h}}\cdot\bm{k}_T)\big)\frac{h_{1Ta}^\perp\bar{h}_{1a}^\perp}{2m_N^3}\bigg]\\
&+\sin(2\phi-\phi_S)\mathcal{F}\bigg[\frac{\hat{\bm{h}}\cdot\bm{p}_T}{m_N}h_{1a}\bar{h}_{1a}^\perp\bigg]\bigg]+\cdots\bigg\},
\end{split}
\end{equation}
where the structure function notation is defined as Eq.
(\ref{Fnotation}), and the azimuthal angles $\phi$ and $\phi_S$ are
defined as shown in Fig.~\ref{angle2}.
Therefore, one can define the following azimuthal asymmetries:
\begin{align}\label{ssas}
&A_{TU}^{\sin(2\phi-\phi_S)}=\frac{\frac{1}{Q^2}\sum_ae_a^2\mathcal{F}[\frac{\hat{\bm{h}}\cdot\bm{p}_T}{m_N}h_{1a}\bar{h}_{1a}^\perp]}
{\frac{1}{Q^2}\sum_ae_a^2\mathcal{F}[f_{1a}\bar{f}_{1a}]},\\
&A_{TU}^{\sin(2\phi+\phi_S)}=\\
&\frac{\frac{1}{Q^2}\sum_ae_a^2\mathcal{F}[\frac{2(\hat{\bm{h}}\cdot\bm{p}_T)
(2(\hat{\bm{h}}\cdot\bm{p}_T)(\hat{\bm{h}}\cdot\bm{k}_T)
-\bm{p}_T\cdot\bm{k}_T)-\bm{p}_T^2(\hat{\bm{h}}\cdot\bm{k}_T)}{2m_N^3}h_{1Ta}^\perp\bar{h}_{1a}^\perp]}
{\frac{1}{Q^2}\sum_ae_a^2\mathcal{F}[f_{1a}\bar{f}_{1a}]},\nonumber\\
&A_{LU}^{\sin2\phi}=\frac{\frac{1}{Q^2}\sum_ae_a^2\mathcal{F}
[\frac{2(\hat{\bm{h}}\cdot\bm{p}_T)(\hat{\bm{h}}\cdot\bm{k}_T)-\bm{p}_T\cdot\bm{k}_T}{m_N^2}h_{1La}^\perp\bar{h}_{1a}^\perp]}
{\frac{1}{Q^2}\sum_ae_a^2\mathcal{F}[f_{1a}\bar{f}_{1a}]}.\label{ssass}
\end{align}

If we take both $\gamma^*$ and $Z$ boson into account, the cross section is expressed as~\cite{Boer:1999mm}
\begin{equation}
\begin{split}
&\frac{d\sigma}{d\Omega dx_1dx_2d^2\bm{q}_T}=\frac{\alpha_{\textrm{em}}}{3Q^2}\sum_a\bigg\{K_1(\theta)\mathcal{F}[f_{1a}\bar{f}_{1a}]\\
&\quad+S_L[K_3(\theta)\sin2\phi+K_4(\theta)\cos2\phi]\\
&\quad\times\mathcal{F}
\bigg[\frac{2(\hat{\bm{h}}\cdot\bm{p}_T)(\hat{\bm{h}}\cdot\bm{k}_T)-\bm{p}_T\cdot\bm{k}_T}{m_N^2}h_{1La}^\perp\bar{h}_{1a}^\perp\bigg]\\
&\quad+|\bm{S}_T|\bigg[[K_3(\theta)\sin(2\phi+\phi_S)+K_4(\theta)\cos(2\phi+\phi_S)]\\
&\quad\times\mathcal{F}\bigg[\big(2(\hat{\bm{h}}\cdot\bm{p}_T)(2(\hat{\bm{h}}\cdot\bm{p}_T)(\hat{\bm{h}}\cdot\bm{k}_T)\\
&\quad\quad-\bm{p}_T\cdot\bm{k}_T)-\bm{p}_T^2(\hat{\bm{h}}\cdot\bm{k}_T)\big)\frac{h_{1Ta}^\perp\bar{h}_{1a}^\perp}{2m_N^3}\bigg]\\
&\quad+[K_3(\theta)\sin(2\phi-\phi_S)+K_4(\theta)\cos(2\phi-\phi_S)]\\
&\quad\times\mathcal{F}\bigg[\frac{\hat{\bm{h}}\cdot\bm{p}_T}{m_N}h_{1a}\bar{h}_{1a}^\perp\bigg]\bigg]+\cdots\bigg\},
\end{split}
\end{equation}
where the coefficients $K_1(\theta)$, $K_3(\theta)$ and $K_4(\theta)$ are defined in (\ref{K1})--(\ref{K4}).
Then the azimuthal asymmetries defined in (\ref{ssas})--(\ref{ssass}) are expressed with $Z$ taken into account as
\begin{align}\label{ssaz}
&A_{TU}^{\sin(2\phi-\phi_S)}=\frac{2\sum_aK_3(\theta)\mathcal{F}[\frac{\hat{\bm{h}}\cdot\bm{p}_T}{m_N}h_{1a}\bar{h}_{1a}^\perp]}
{\sum_aK_1(\theta)\mathcal{F}[f_{1a}\bar{f}_{1a}]},\\
&A_{TU}^{\sin(2\phi+\phi_S)}=\\
&\frac{2\sum_aK_3(\theta)\mathcal{F}[\frac{2(\hat{\bm{h}}\cdot\bm{p}_T)
(2(\hat{\bm{h}}\cdot\bm{p}_T)(\hat{\bm{h}}\cdot\bm{k}_T)
-\bm{p}_T\cdot\bm{k}_T)-\bm{p}_T^2(\hat{\bm{h}}\cdot\bm{k}_T)}{2m_N^3}h_{1Ta}^\perp\bar{h}_{1a}^\perp]}
{\sum_aK_1(\theta)\mathcal{F}[f_{1a}\bar{f}_{1a}]},\nonumber\\
&A_{LU}^{\sin2\phi}=\frac{2\sum_aK_3(\theta)\mathcal{F}
[\frac{2(\hat{\bm{h}}\cdot\bm{p}_T)(\hat{\bm{h}}\cdot\bm{k}_T)-\bm{p}_T\cdot\bm{k}_T}{m_N^2}h_{1La}^\perp\bar{h}_{1a}^\perp]}
{\sum_aK_1(\theta)\mathcal{F}[f_{1a}\bar{f}_{1a}]}.
\end{align}
The $\cos2\phi$, $\cos(2\phi-\phi_S)$ and $\cos(2\phi+\phi_S)$ azimuthal dependent terms are $1/Q^2$
suppressed. This suppression can be found from (\ref{K4}) and (\ref{chi3}).

For $pd$ dilepton production processes with the deuteron
longitudinally or transversely polarized, we assume the isospin
symmetry and neglect the nuclear effect as we do for unpolarized
$pd$ processes. Therefore, we can get the $\sin(2\phi-\phi_S)$,
$\sin(2\phi+\phi_S)$ and $\sin2\phi$ azimuthal asymmetries of $pd$
dilepton production processes in Drell--Yan continuum region and around the $Z$ pole by replacing
the distribution functions of the target via
(\ref{replacingdeutron}).

\section{Numerical results}

In this section, we calculate the azimuthal asymmetries of the $pp$
and $pd$ dilepton production processes with the proton or deuteron
target unpolarized and longitudinally or transversally polarized in
Drell--Yan continuum region and $Z$ resonance
region respectively. We present a numerical estimation of these
azimuthal asymmetries for measurement at a fixed-target experiment
using the LHC beams proposed by Brodsky, Fleuret, Hadjidakis and
Lansberg~\cite{Brodsky:2012vg}. With the $7$ TeV proton beams, the
center-of-mass frame energy $\sqrt{s}=115$ GeV for two nucleons.

The cross section of Drell--Yan process can also be expressed depending on $y$ and $Q^2$ instead of $x_1$ and $x_2$
with just a Jacobian multiplied as
\begin{equation}
\frac{d\sigma}{dydQ^2d^2\bm{q}_Td\Omega}=\frac{1}{s}\frac{d\sigma}{dx_1dx_2d^2\bm{q}_Td\Omega}.
\end{equation}
At the region $\bm{q}_T^2\ll Q^2$, we have the following relation:
\begin{equation}
x_1=\frac{Q}{\sqrt{s}}e^y,\quad x_2=\frac{Q}{\sqrt{s}}e^{-y}.
\end{equation}
The $x_1$ and $x_2$ can also be expressed with $x_F$ and $Q^2$
\begin{equation}
\begin{split}
x_1=\frac{1}{2}\left(x_F+\sqrt{x_F^2+4\frac{Q^2}{s}}\right),\\
x_2=\frac{1}{2}\left(-x_F+\sqrt{x_F^2+4\frac{Q^2}{s}}\right),
\end{split}
\end{equation}
and
\begin{equation}
x_F=x_1-x_2.
\end{equation}
$x_1$ is the momentum fraction of the parton in the beam proton,
and $x_2$ is that of the parton in the target nucleon.
For the single polarized processes, $x_2$ is the momentum fraction of the parton in the polarized target nucleon,
sometimes labeled as $x^\uparrow$ in the literature.
To calculate the azimuthal asymmetries depending on $Q$, $x_F$ or $q_T$, we need to integrate over
the other variables in the numerator and the dominator of the expression of the asymmetries respectively.
The rapidity is cut in $[-4.8,1]$ which is the easiest region to carry on measurements as discussed in~\cite{Brodsky:2012vg}, and this is where the momentum fraction of the parton inside the polarized nucleons is the largest.

In our calculation, we adopt the Boer--Mulders function $h_1^\perp$ extracted from the
unpolarized $pd$ and $pp$ Drell--Yan data~\cite{Zhu:2006gx,Zhu:2008sj,Zhang:2008nu,Lu:2009ip}.
The parametrization of $h_1^\perp$ for both valence and sea quarks has the form:
\begin{equation}
h_{1q}^\perp(x, k_T^2)=H_qx^{c^q}(1-x)^{b}f_{1q}(x)\frac{1}{\pi k_{\textrm{bm}}^2}\exp\bigg(\frac{-\bm{k}_T^2}{k_{\textrm{bm}}^2}\bigg),
\end{equation}
where the subscript "bm" stands for the Boer--Mulders functions, and $q=u$, $d$, $\bar{u}$ and $\bar{d}$.
The possible range of the parameters $H_q$ allowed by the positivity bound~\cite{Bacchetta:1999kz} can
be described by a coefficient $\omega$ which balance the contributions of quark and antiquark. $H_q\rightarrow\omega H_q$ for $q=u$, $d$ and $H_q\rightarrow\omega^{-1}H_q$ for $q=\bar{u}$, $\bar{d}$ will not change the calculated $\cos2\phi$ asymmetry in the unpolarized $pp$ and $pd$ Drell--Yan data.
The range of $\omega$ is $0.48<\omega<2.1$, and we choose the case $\omega=1$, which corresponds to the central values of $H_q$, in our calculation.

\begin{figure}
  \includegraphics[width=0.5\textwidth]{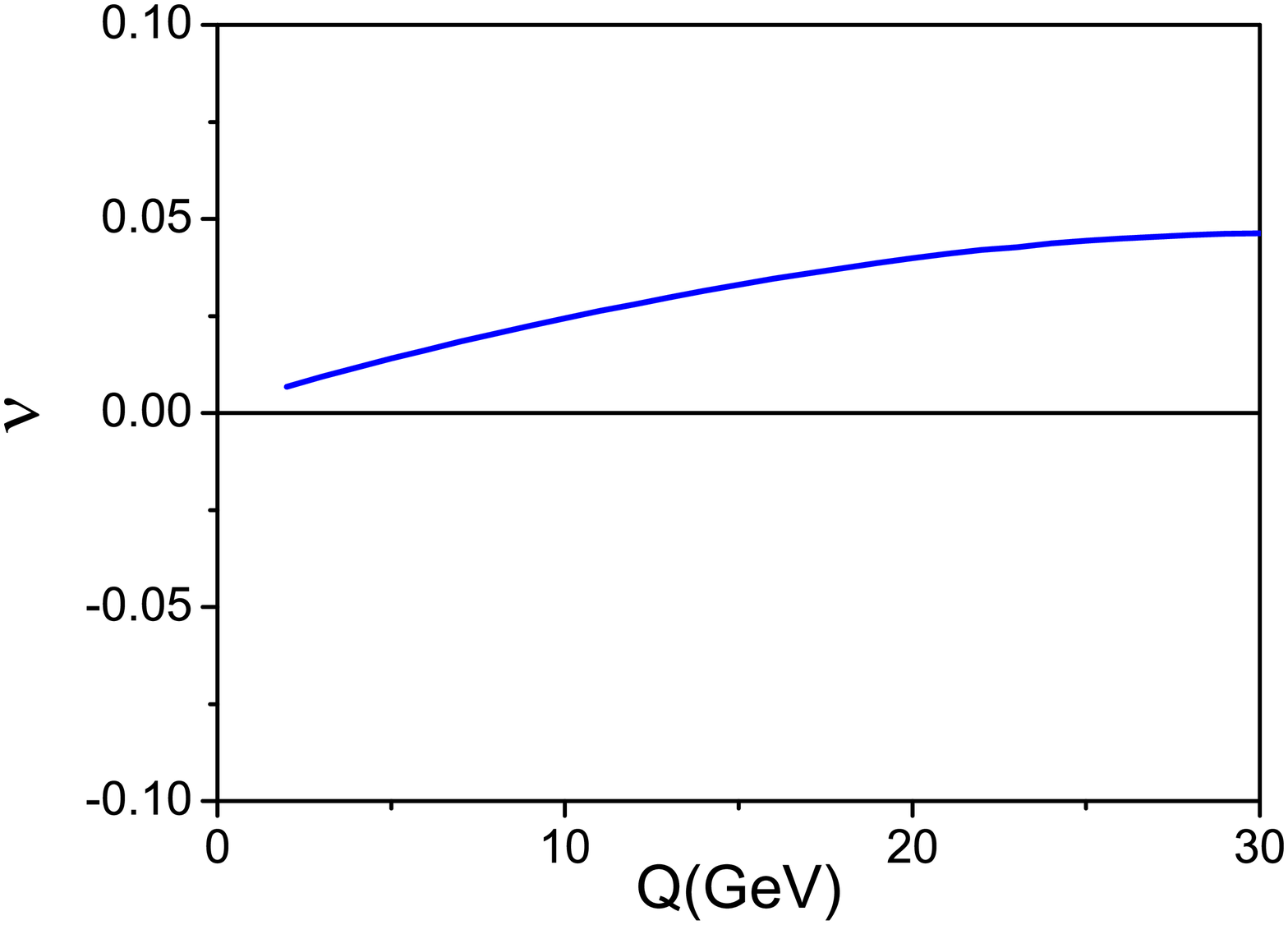}
\caption{The $\cos2\phi$ azimuthal asymmetry depending on $Q$ of unpolarized $pp$ Drell--Yan process with both $\gamma^*$ and $Z$ taken into account and allowed rapidity integrated in the cut $[-4.8, 1]$. The same cut of rapidity is chosen in Figs.\ref{unpdcont1},\ref{sppcont1a}-\ref{spdcont1c}.}
\label{unppcont1}       
  \includegraphics[width=0.5\textwidth]{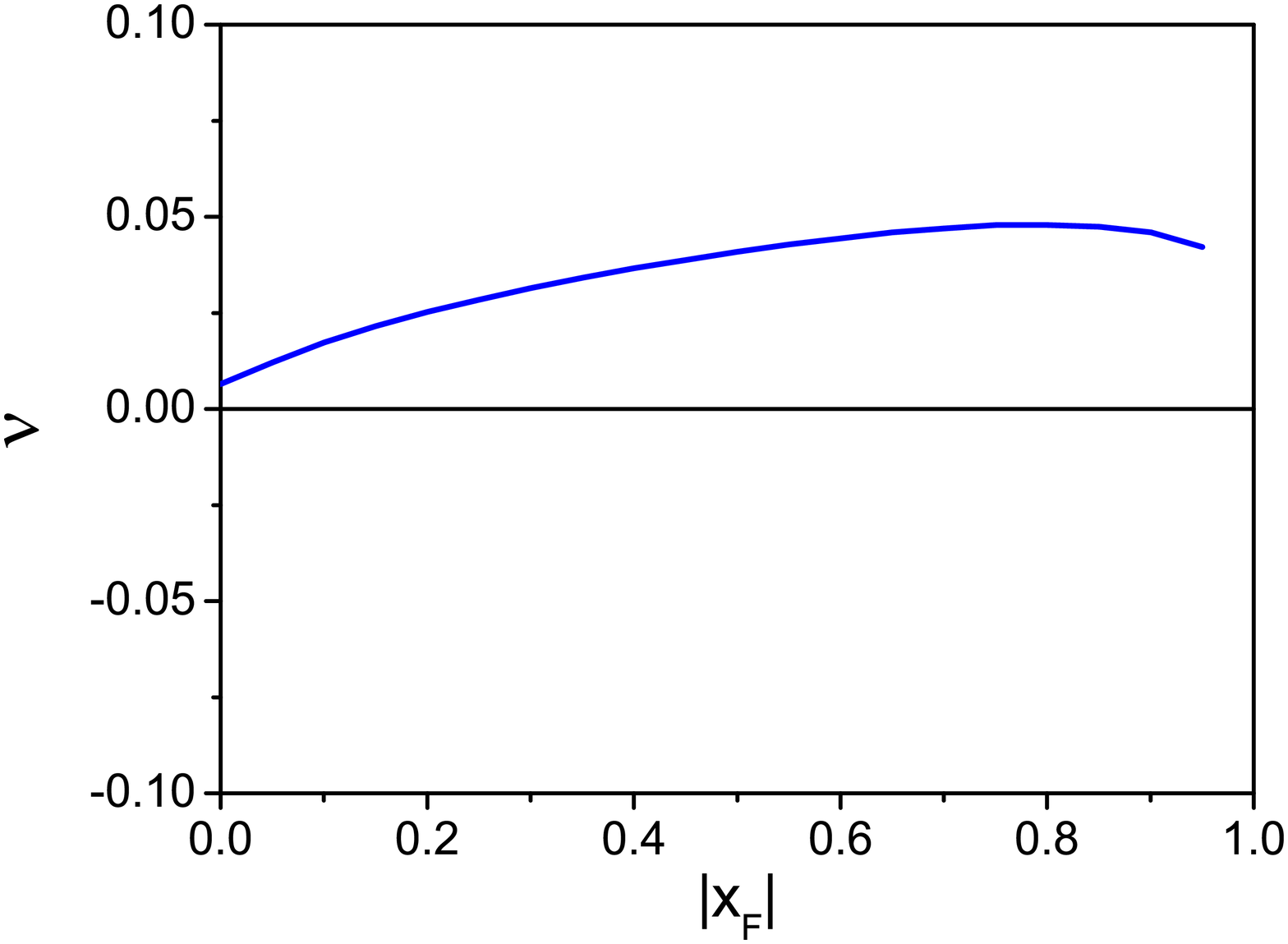}
\caption{The $\cos2\phi$ azimuthal asymmetry depending on $x_F$ of unpolarized $pp$ Drell--Yan process at $Q=2$ GeV.}
\label{unppcontxf2}       
\end{figure}

\begin{figure}
  \includegraphics[width=0.5\textwidth]{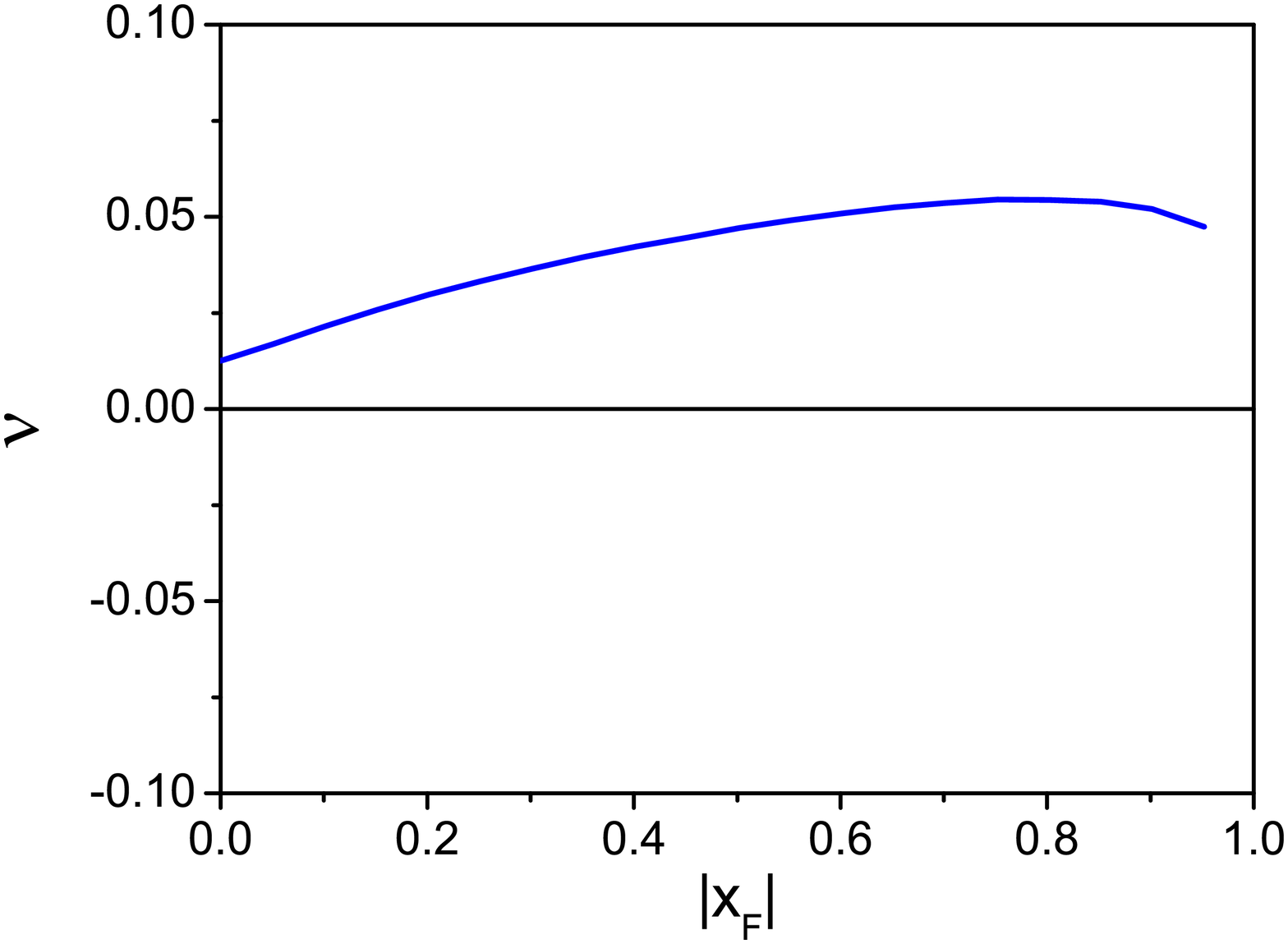}
\caption{The $\cos2\phi$ azimuthal asymmetry depending on $x_F$ of unpolarized $pp$ Drell--Yan process at $Q=5$ GeV.}
\label{unppcontxf5}       
  \includegraphics[width=0.5\textwidth]{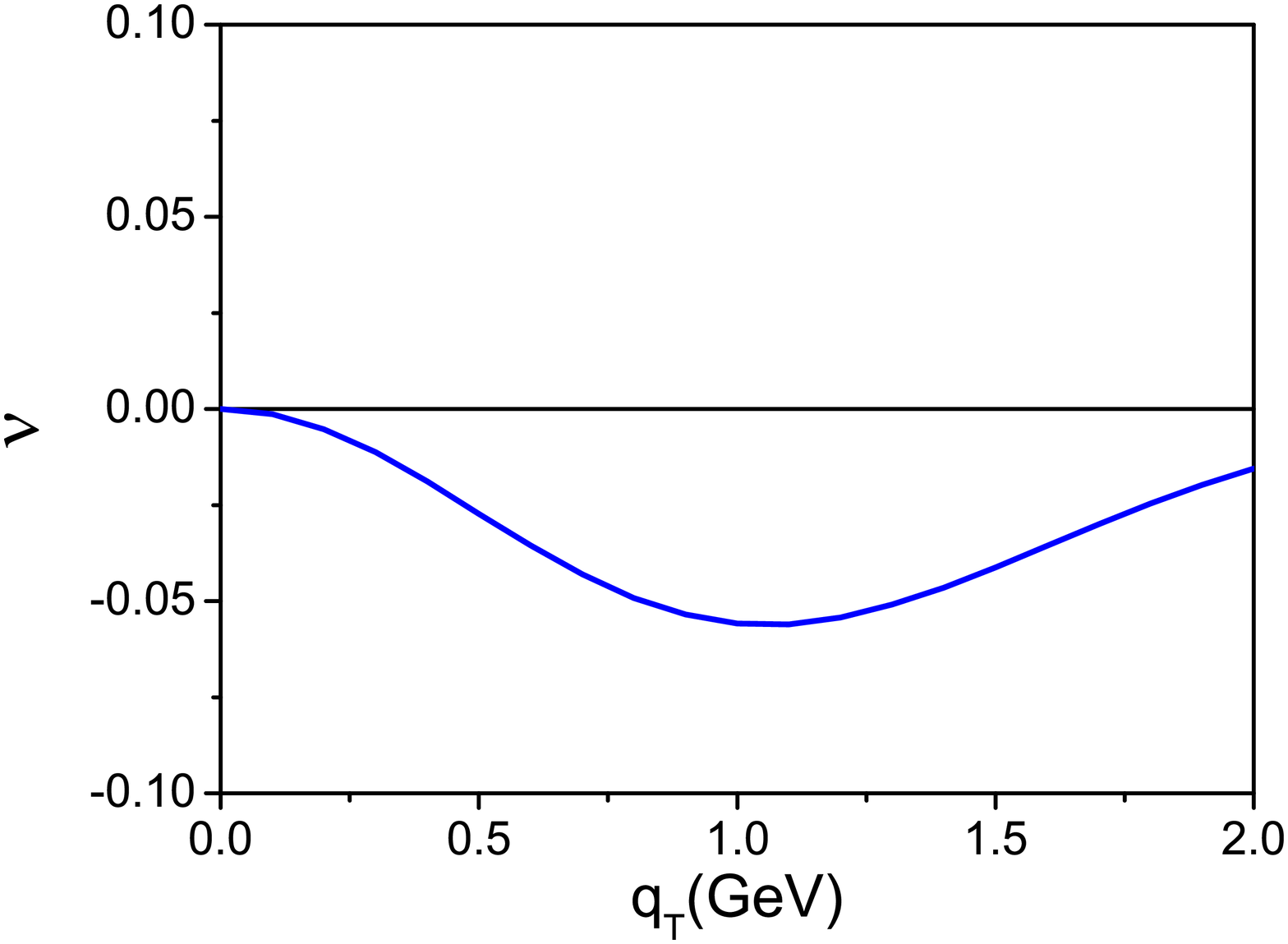}
\caption{The $\cos2\phi$ azimuthal asymmetry depending on $q_T$ of unpolarized $pp$ process in $Z$ resonance region.}
\label{unppz}       
\end{figure}

\begin{figure}
  \includegraphics[width=0.5\textwidth]{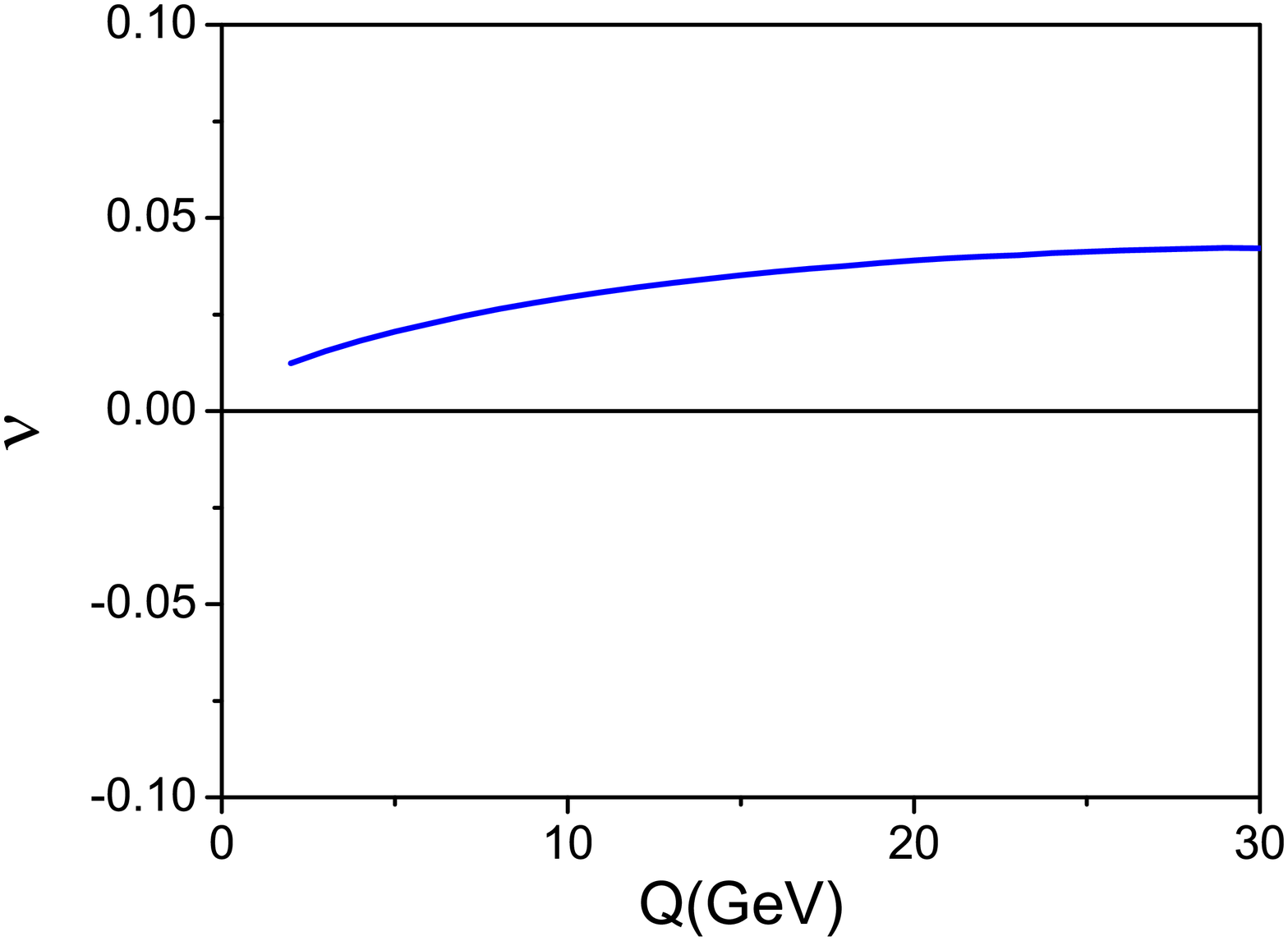}
\caption{The $\cos2\phi$ azimuthal asymmetry depending on $Q$ of unpolarized $pd$ Drell--Yan process with both $\gamma^*$ and $Z$ taken into account.}
\label{unpdcont1}       
  \includegraphics[width=0.5\textwidth]{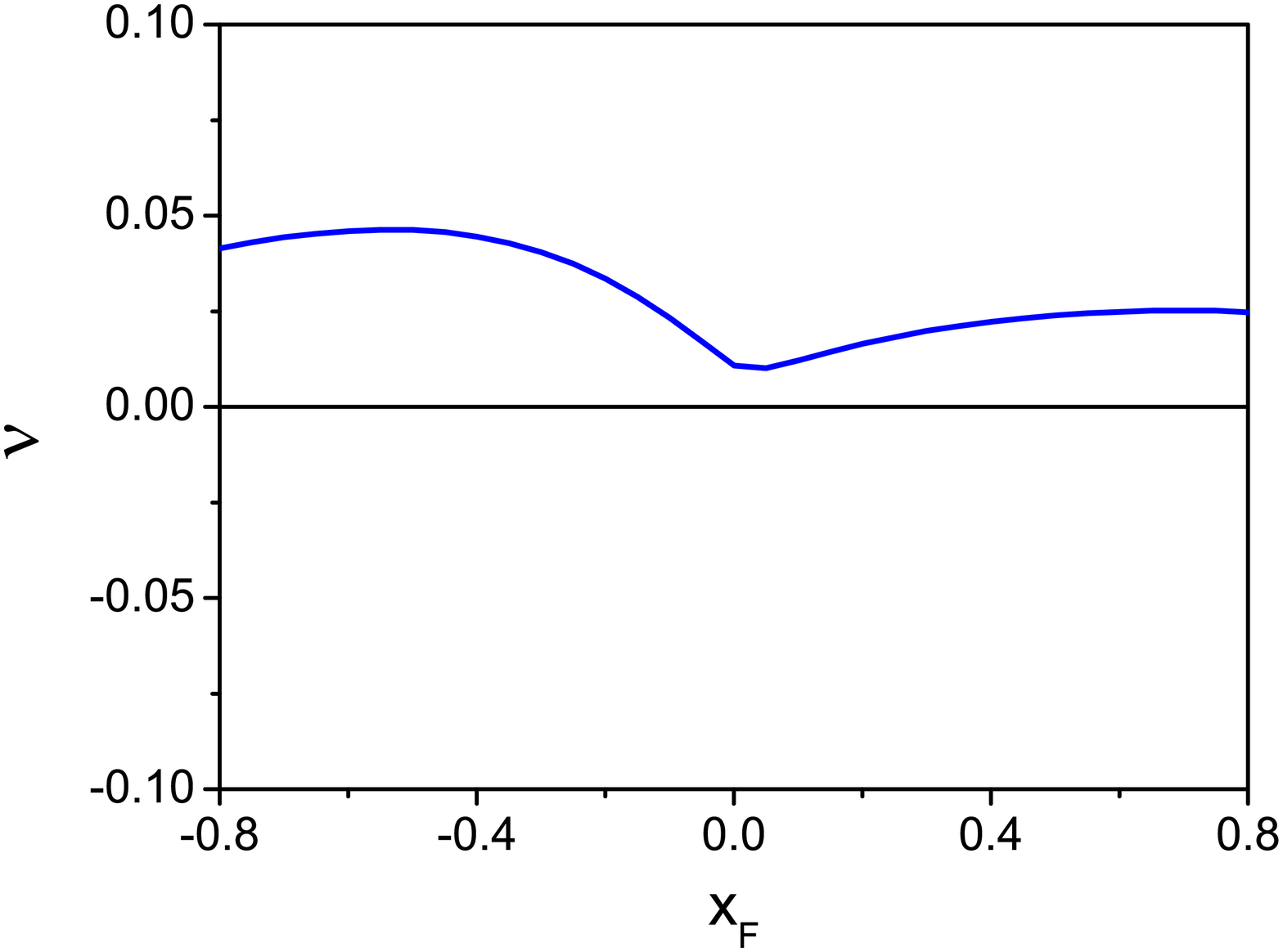}
\caption{The $\cos2\phi$ azimuthal asymmetry depending on $x_F$ of unpolarized $pd$ Drell--Yan process at $Q=2$ GeV.}
\label{unpdcontxf2}       
\end{figure}

\begin{figure}
  \includegraphics[width=0.5\textwidth]{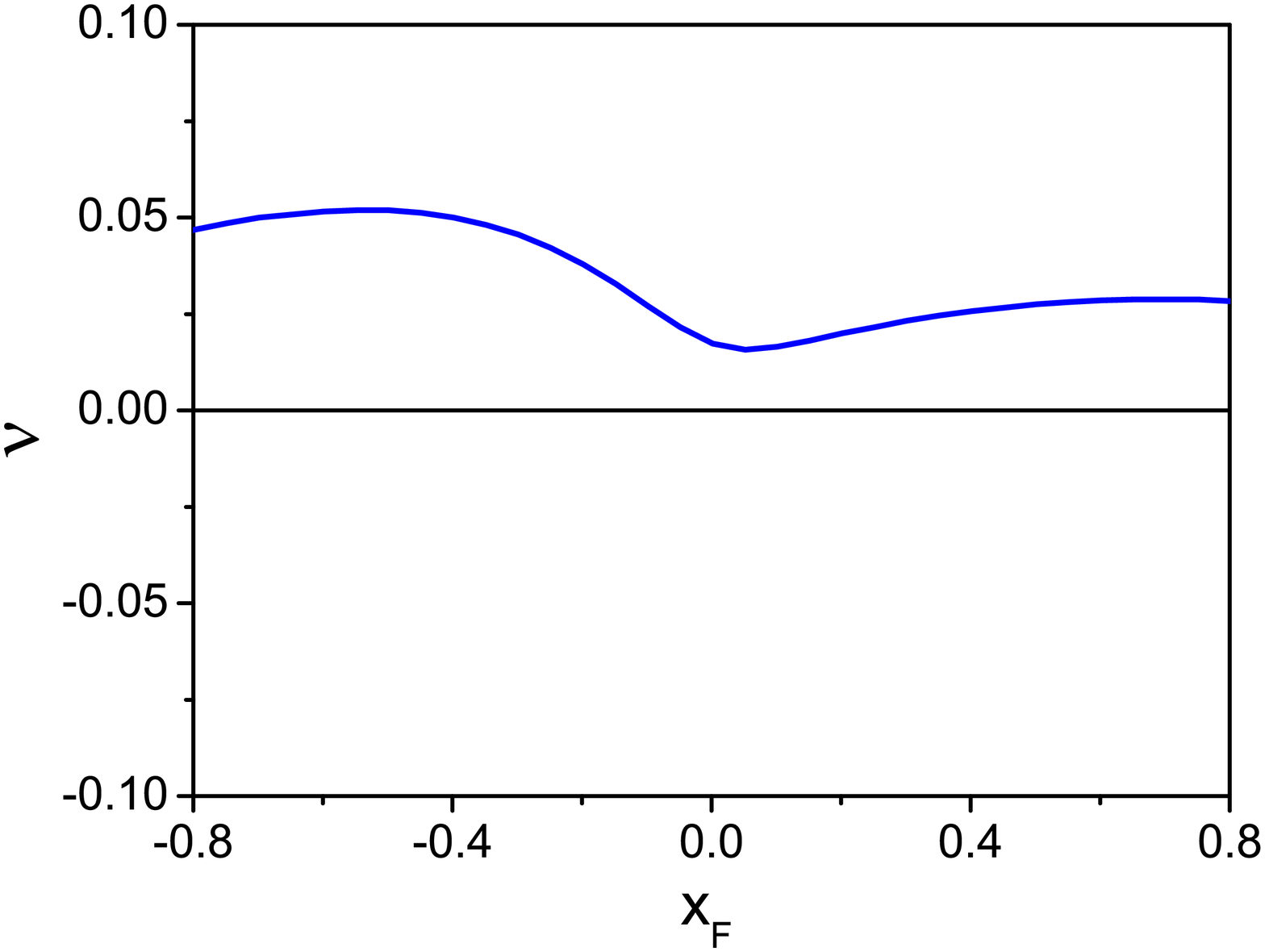}
\caption{The $\cos2\phi$ azimuthal asymmetry depending on $x_F$ of unpolarized $pd$ Drell--Yan process at $Q=5$ GeV.}
\label{unpdcontxf5}       
  \includegraphics[width=0.5\textwidth]{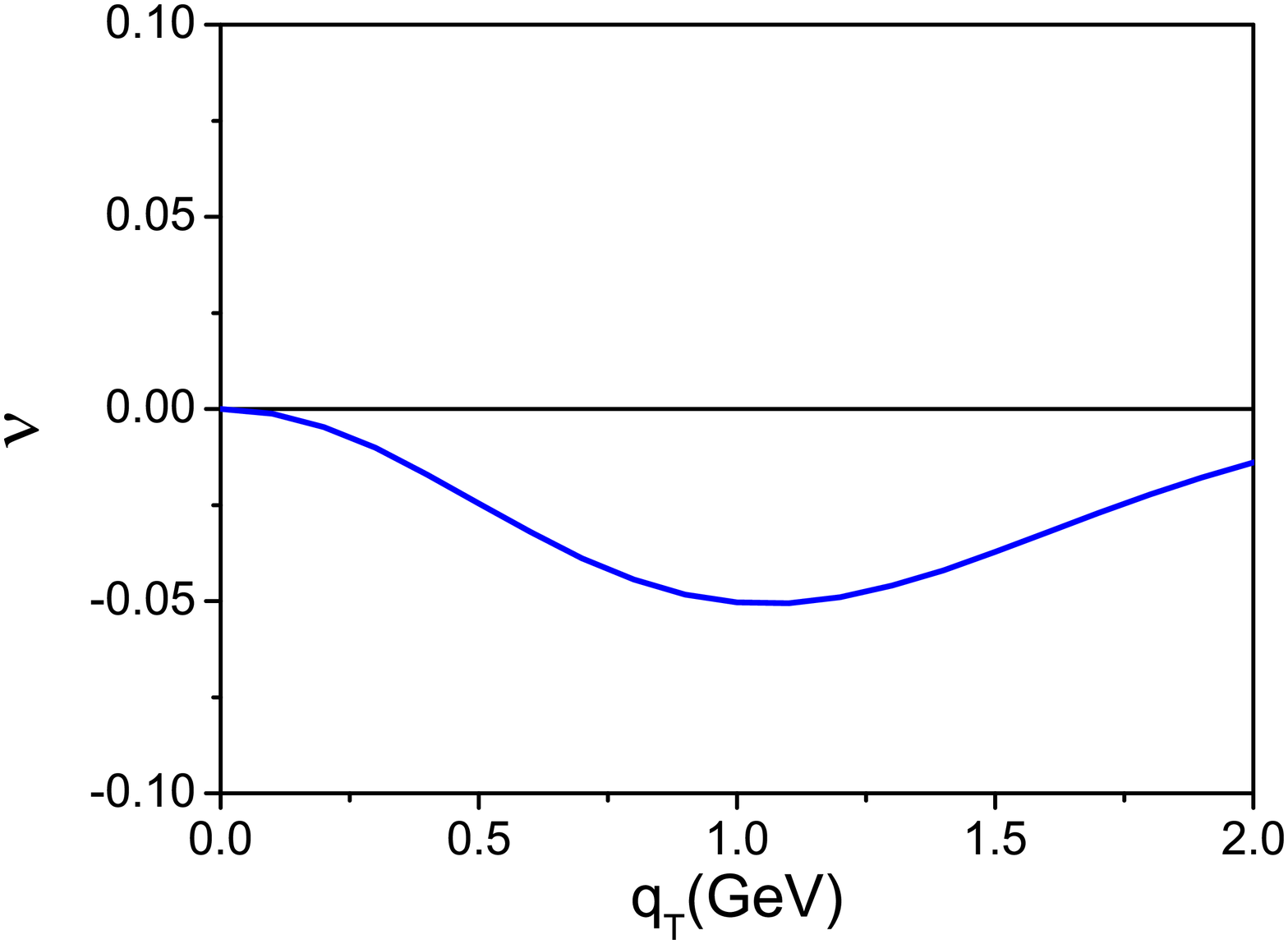}
\caption{The $\cos2\phi$ azimuthal asymmetry depending on $q_T$ of unpolarized $pd$ process in $Z$ resonance region.}
\label{unpdz}       
\end{figure}

In Figs.~\ref{unppcont1} and ~\ref{unpdcont1}, we show the $\cos2\phi$ azimuthal asymmetry
depending on $Q$ from $2$ GeV to $30$ GeV
of the unpolarized $pp$ and $pd$ Drell--Yan process at AFTER including $Z$
taken into account. Figs.~\ref{unppcontxf2},~\ref{unppcontxf5},~\ref{unpdcontxf2} and~\ref{unpdcontxf5}
respectively show the $\cos2\phi$ azimuthal
asymmetry depending on $x_F$ of the unpolarized $pp$ and $pd$ processes with $Q=2$ GeV and $Q=5$ GeV
as for low and mid $Q$ Drell--Yan regions at AFTER. Figs.~\ref{unppz} and~\ref{unpdz} respectively show
this azimuthal asymmetry of $pp$ and $pd$ processes at the $Z$ pole at AFTER.


\begin{figure}
  \includegraphics[width=0.5\textwidth]{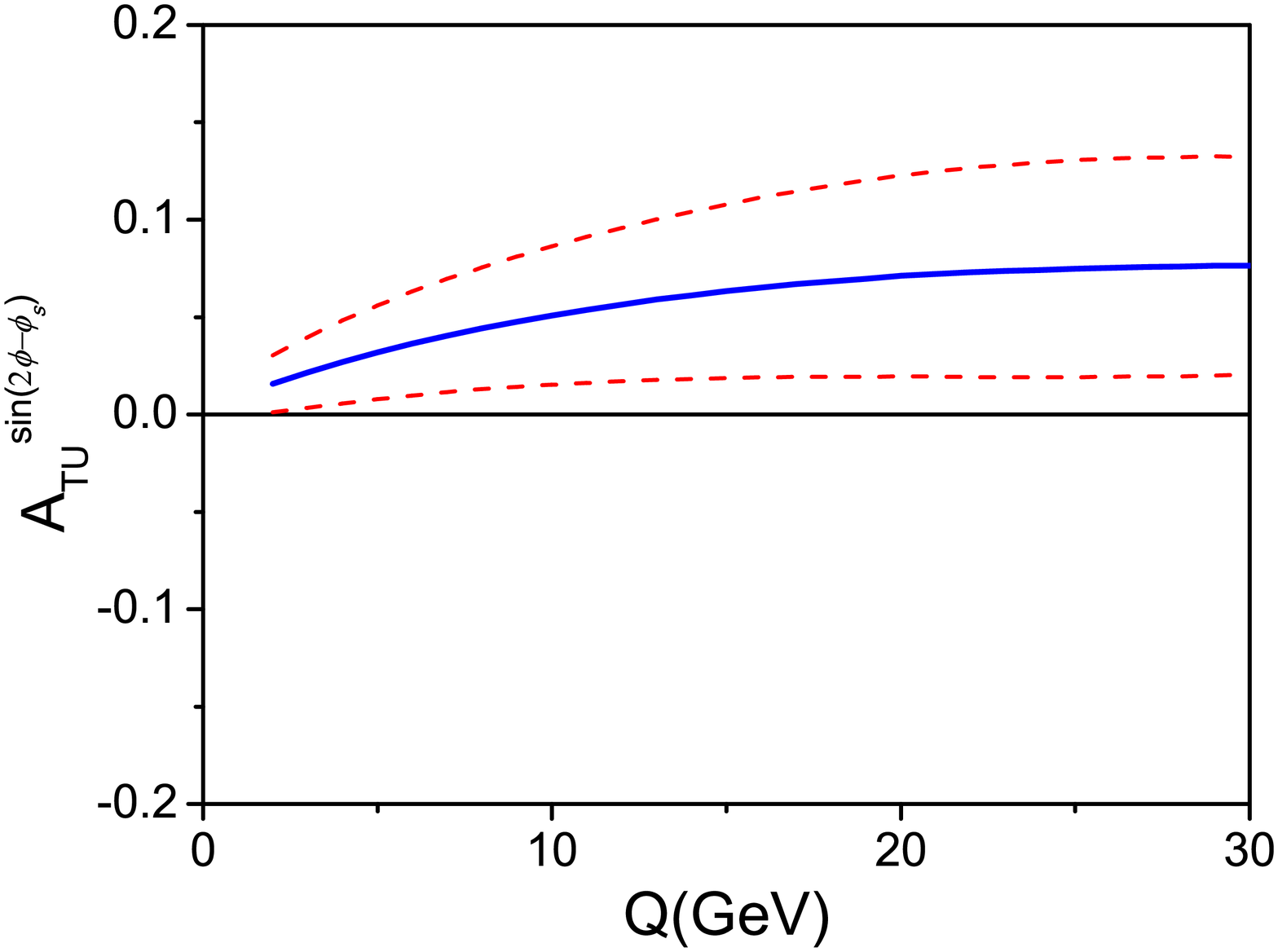}
\caption{The $\sin(2\phi-\phi_S)$ azimuthal asymmetry
$A_{TU}^{\sin(2\phi-\phi_S)}$ depending on $Q$ of target proton
polarized $pp$ Drell--Yan process with both $\gamma^*$ and $Z$ taken
into account. The dashed curves show the range of the asymmetry by
considering the additional distributions of sea quarks constrained
by the positivity bounds, corresponding to the same case as Figs.~\ref{sppcont2b}--\ref{spdzc}.}
\label{sppcont1a}       
  \includegraphics[width=0.5\textwidth]{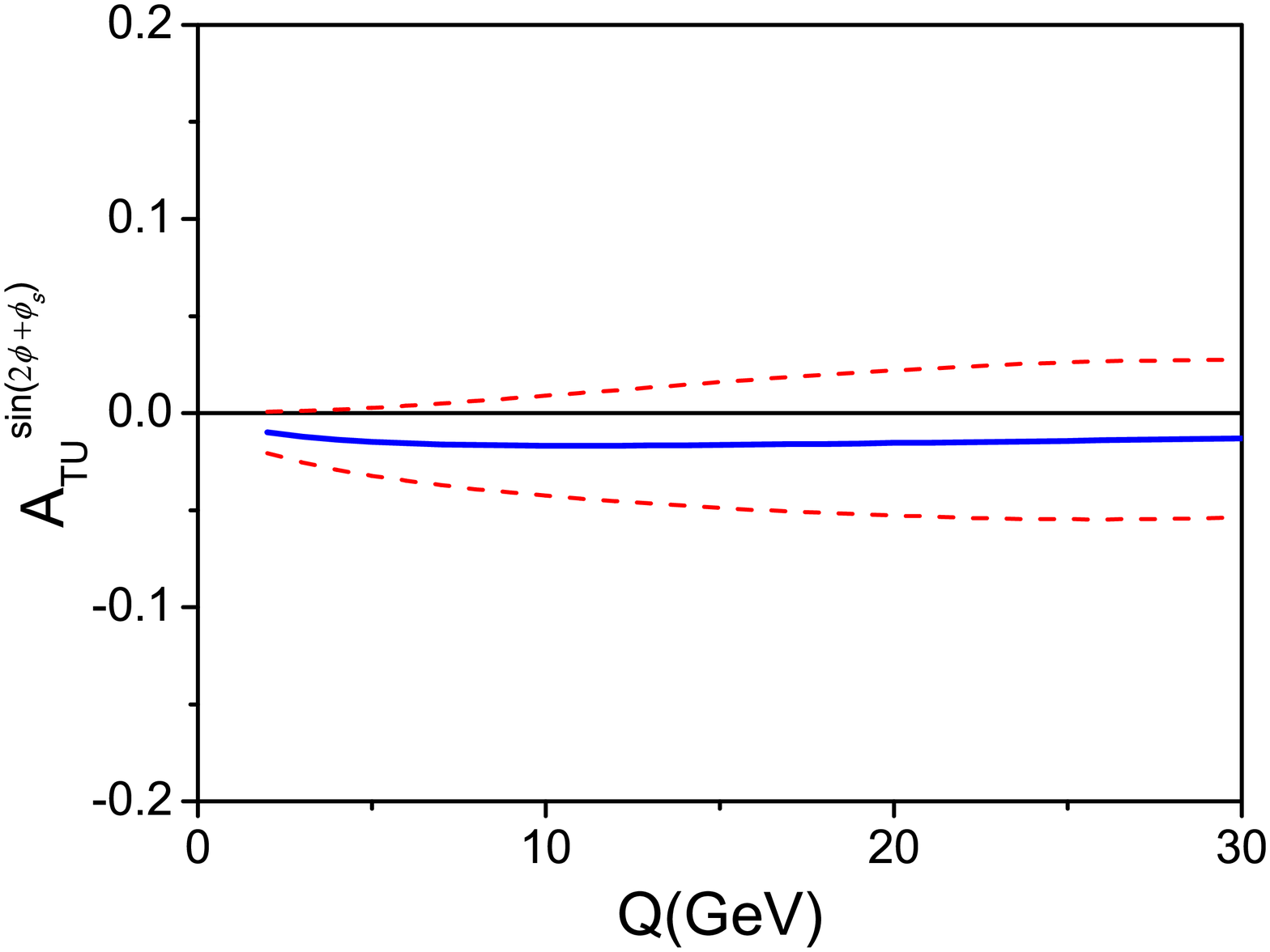}
\caption{The $\sin(2\phi+\phi_S)$ azimuthal asymmetry
$A_{TU}^{\sin(2\phi+\phi_S)}$ depending on $Q$ of target proton
polarized $pp$ Drell--Yan process with both $\gamma^*$ and $Z$ taken
into account.}
\label{sppcont1b}       
  \includegraphics[width=0.5\textwidth]{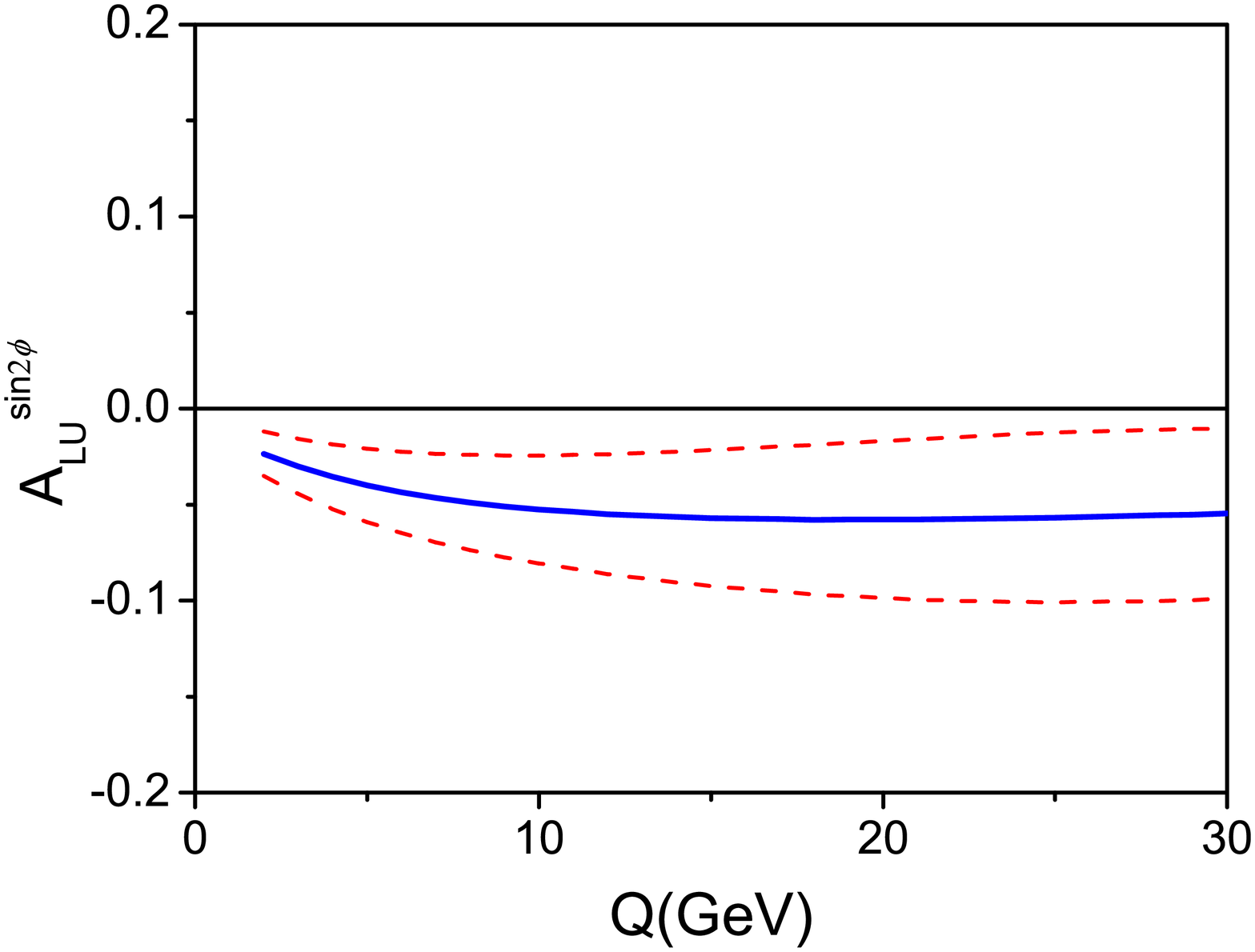}
\caption{The $\sin2\phi$ azimuthal asymmetry $A_{LU}^{\sin2\phi}$ depending on $Q$ of target proton polarized $pp$ Drell--Yan process with both $\gamma^*$ and $Z$ taken into account.}
\label{sppcont1c}       
\end{figure}
\begin{figure}
  \includegraphics[width=0.5\textwidth]{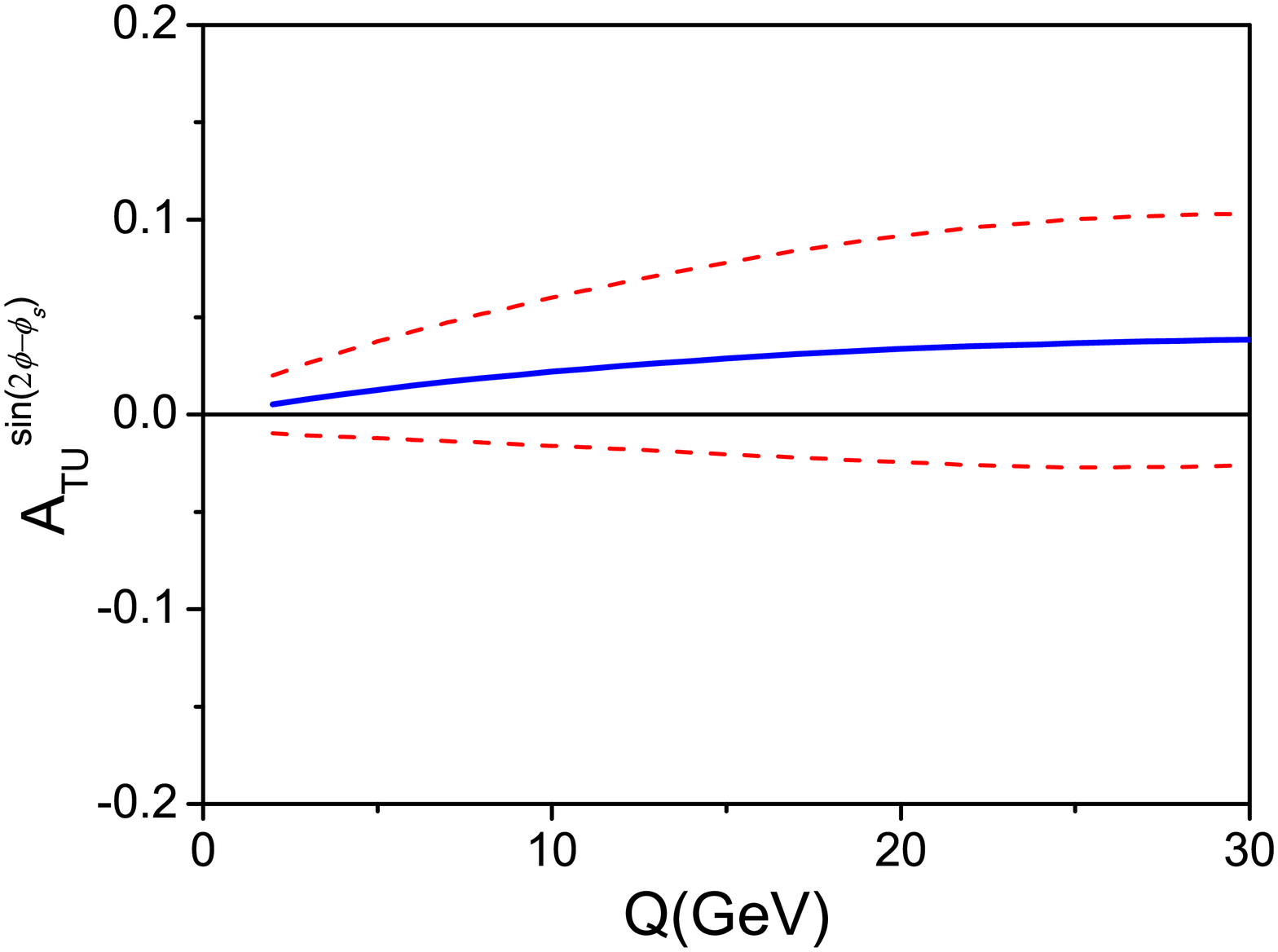}
\caption{The $\sin(2\phi-\phi_S)$ azimuthal asymmetry
$A_{TU}^{\sin(2\phi-\phi_S)}$ depending on $Q$ of target deuteron
polarized $pd$ Drell--Yan process with both $\gamma^*$ and $Z$ taken
into account.}
\label{spdcont1a}       
  \includegraphics[width=0.5\textwidth]{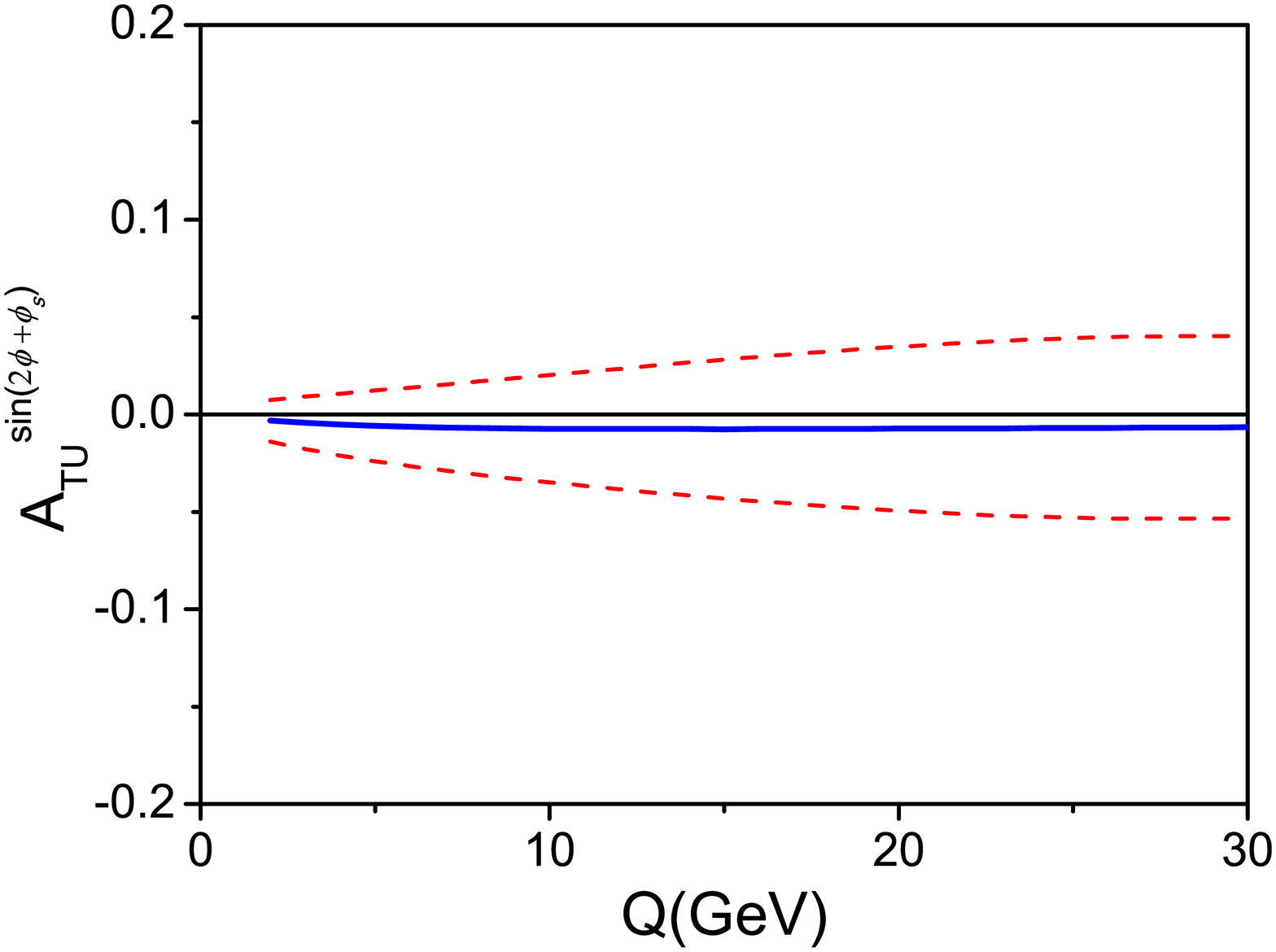}
\caption{The $\sin(2\phi+\phi_S)$ azimuthal asymmetry
$A_{TU}^{\sin(2\phi+\phi_S)}$ depending on $Q$ of target deuteron
polarized $pd$ Drell--Yan process with both $\gamma^*$ and $Z$ taken
into account.}
\label{spdcont1b}       
  \includegraphics[width=0.5\textwidth]{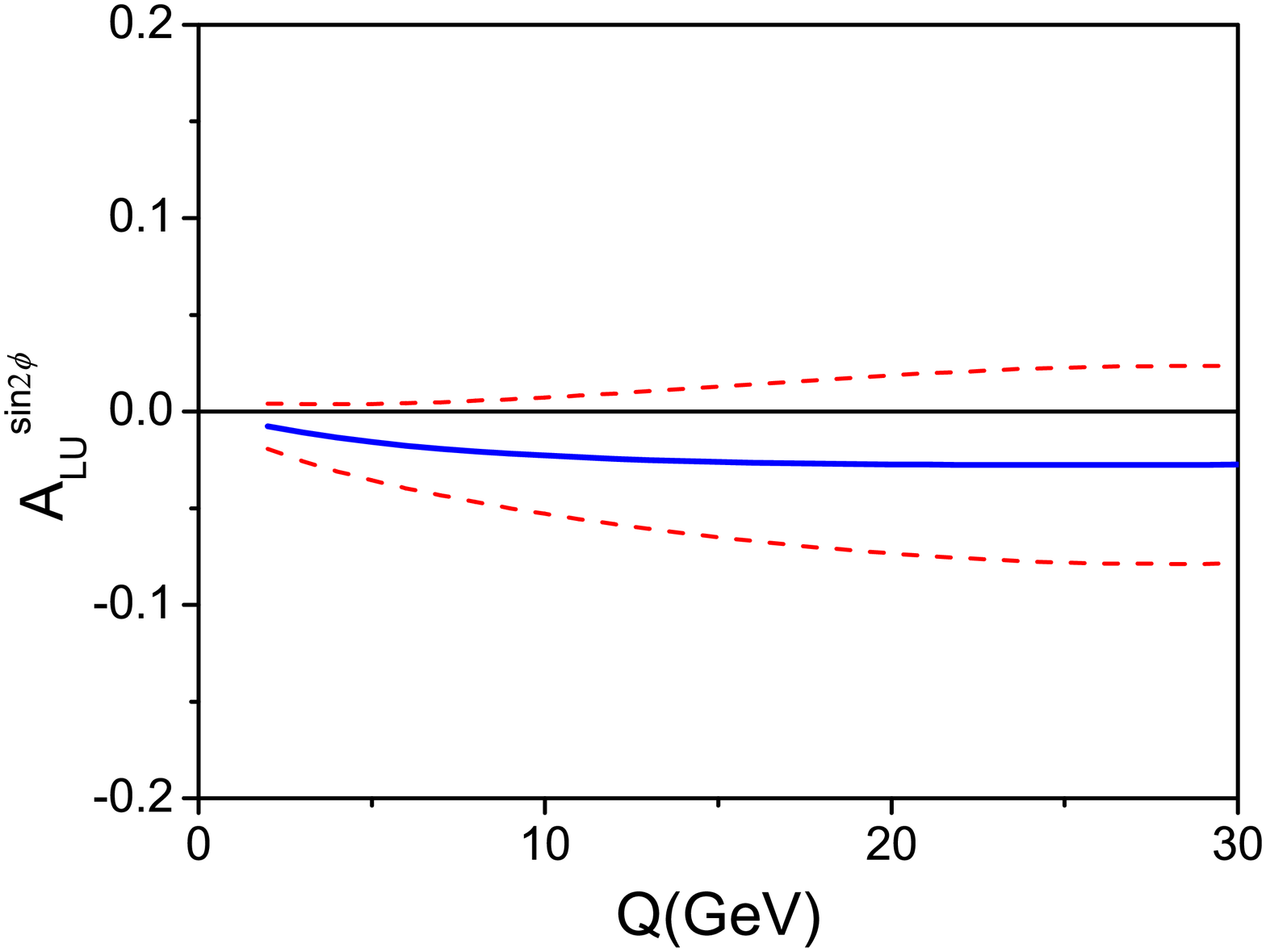}
\caption{The $\sin2\phi$ azimuthal asymmetry $A_{LU}^{\sin2\phi}$ depending on $Q$ of target deuteron
polarized $pd$ Drell--Yan process with both $\gamma^*$ and $Z$ taken into account.}
\label{spdcont1c}       
\end{figure}

\begin{figure}
  \includegraphics[width=0.5\textwidth]{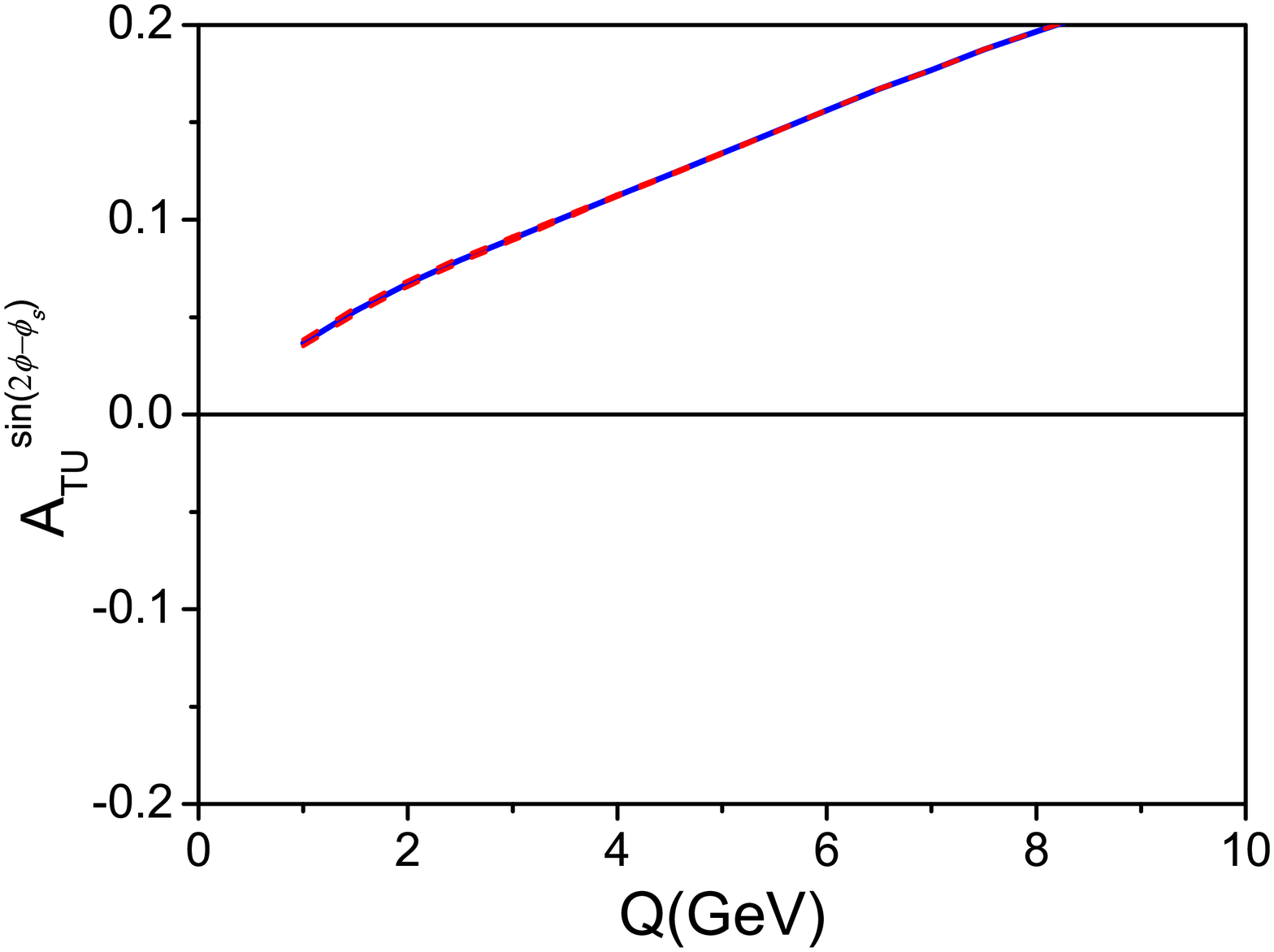}
\caption{The $\sin(2\phi-\phi_S)$ azimuthal asymmetry
$A_{TU}^{\sin(2\phi-\phi_S)}$ depending on $Q$ of target proton
polarized $pp$ Drell--Yan process with both $\gamma^*$ and $Z$ taken
into account and allowed rapidity integrated in the cut $[-4.8, -2]$. The same cut of rapidity is chosen in Figs.\ref{sppcont2b}-\ref{spdcont2c}.}
\label{sppcont2a}       
  \includegraphics[width=0.5\textwidth]{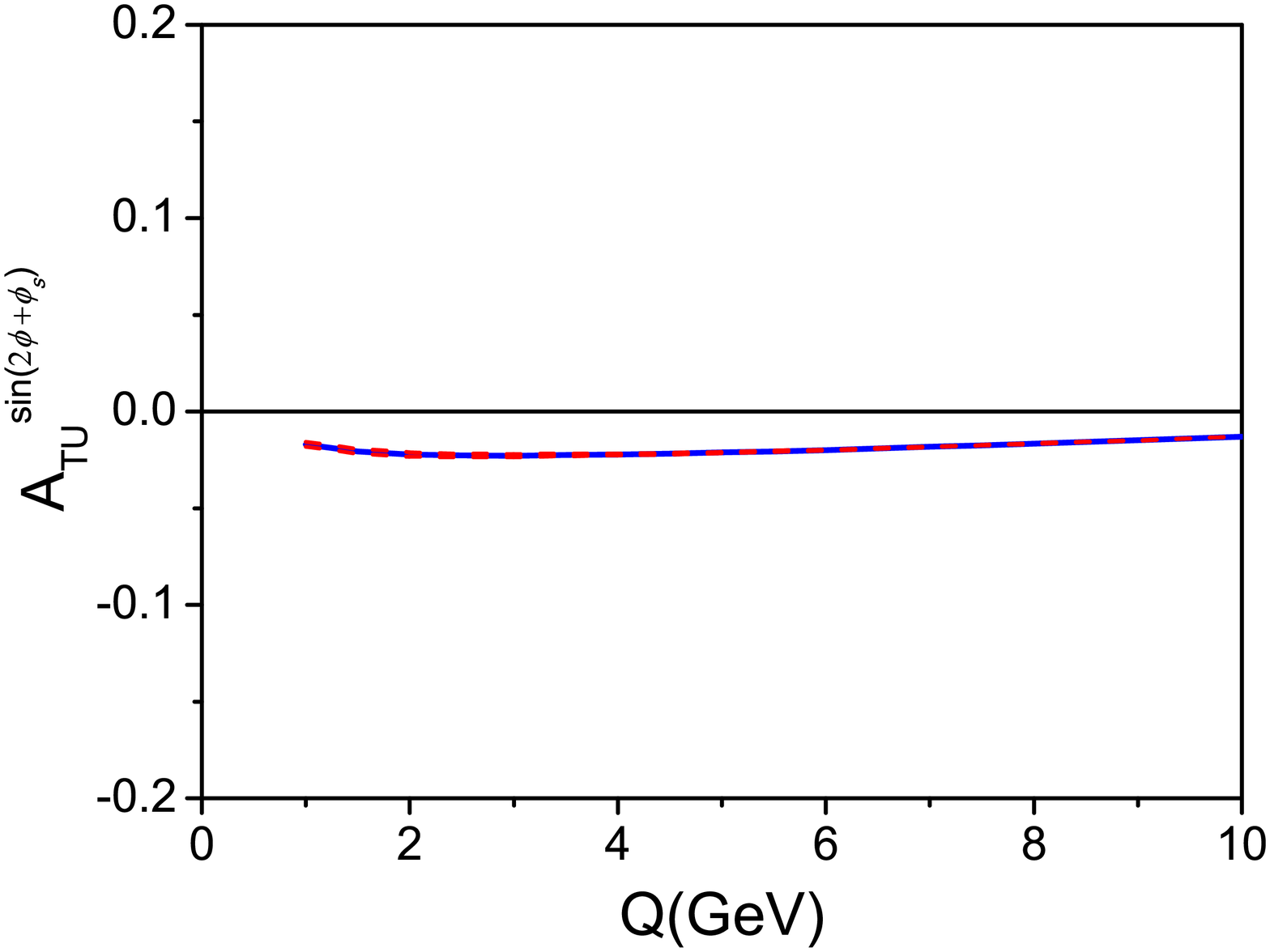}
\caption{The $\sin(2\phi+\phi_S)$ azimuthal asymmetry
$A_{TU}^{\sin(2\phi+\phi_S)}$ depending on $Q$ of target proton
polarized $pp$ Drell--Yan process with both $\gamma^*$ and $Z$ taken
into account.}
\label{sppcont2b}       
  \includegraphics[width=0.5\textwidth]{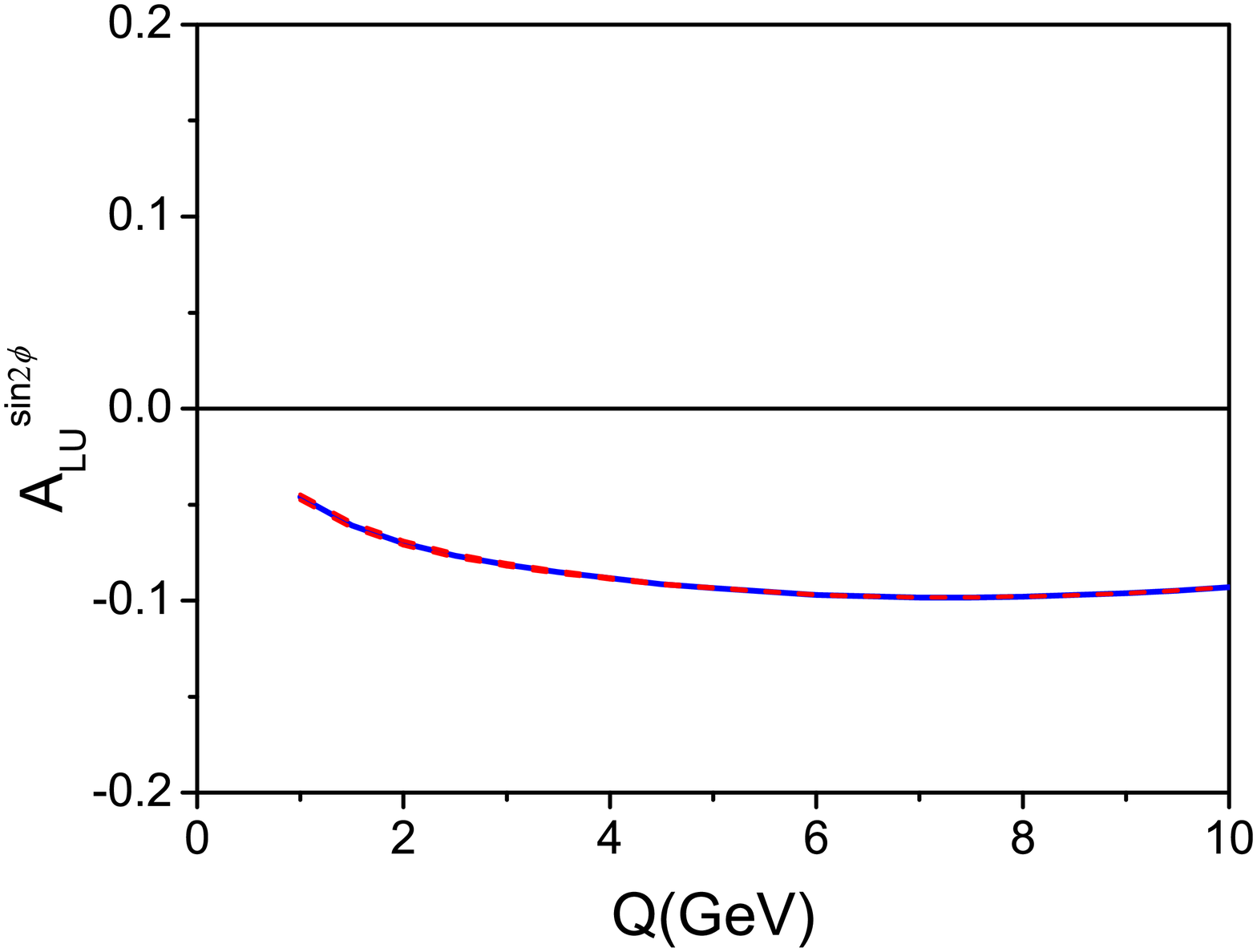}
\caption{The $\sin2\phi$ azimuthal asymmetry $A_{LU}^{\sin2\phi}$ depending on $Q$ of target proton polarized $pp$ Drell--Yan process with both $\gamma^*$ and $Z$ taken into account.}
\label{sppcont2c}       
\end{figure}
\begin{figure}
  \includegraphics[width=0.5\textwidth]{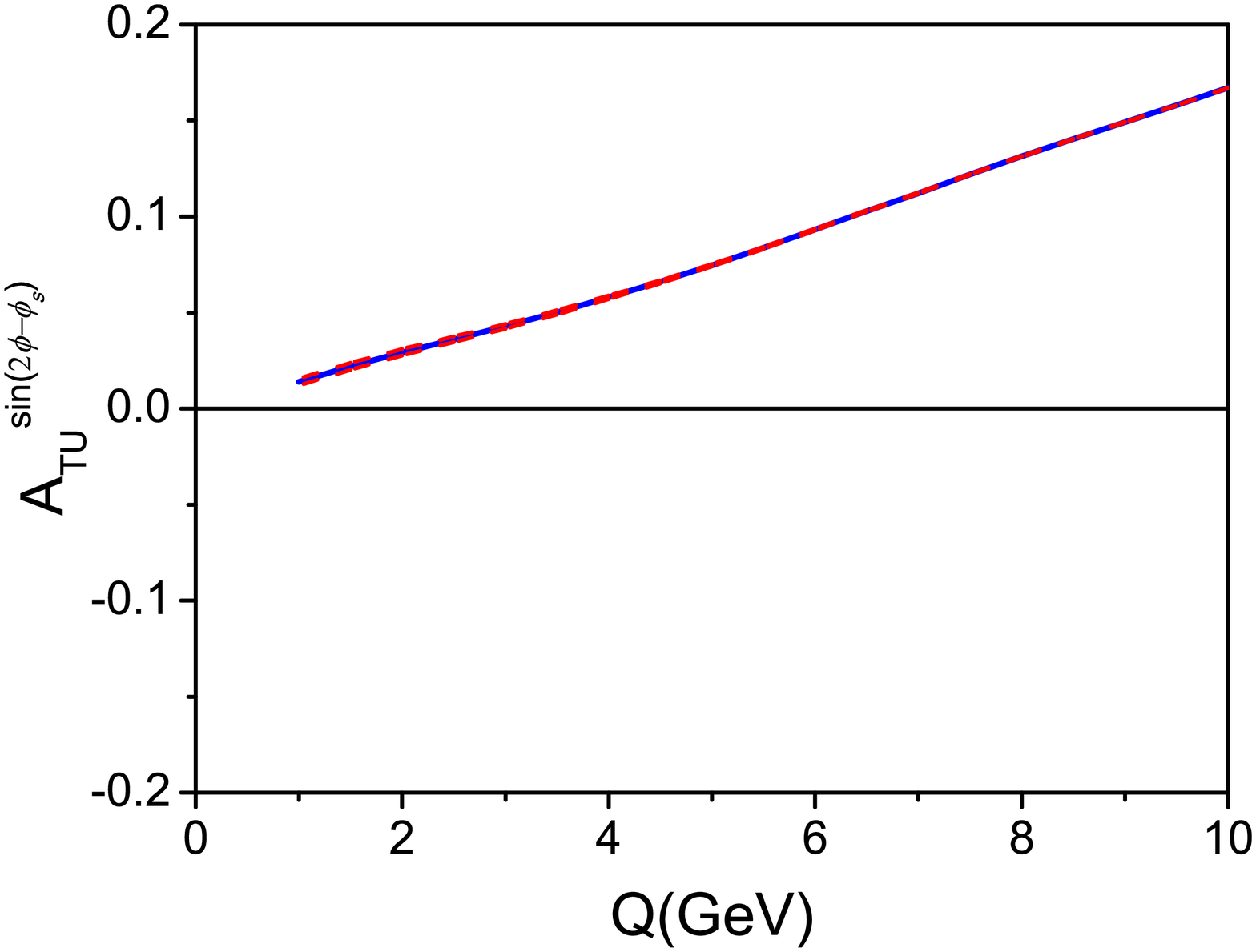}
\caption{The $\sin(2\phi-\phi_S)$ azimuthal asymmetry
$A_{TU}^{\sin(2\phi-\phi_S)}$ depending on $Q$ of target deuteron
polarized $pd$ Drell--Yan process with both $\gamma^*$ and $Z$ taken
into account.}
\label{spdcont2a}       
  \includegraphics[width=0.5\textwidth]{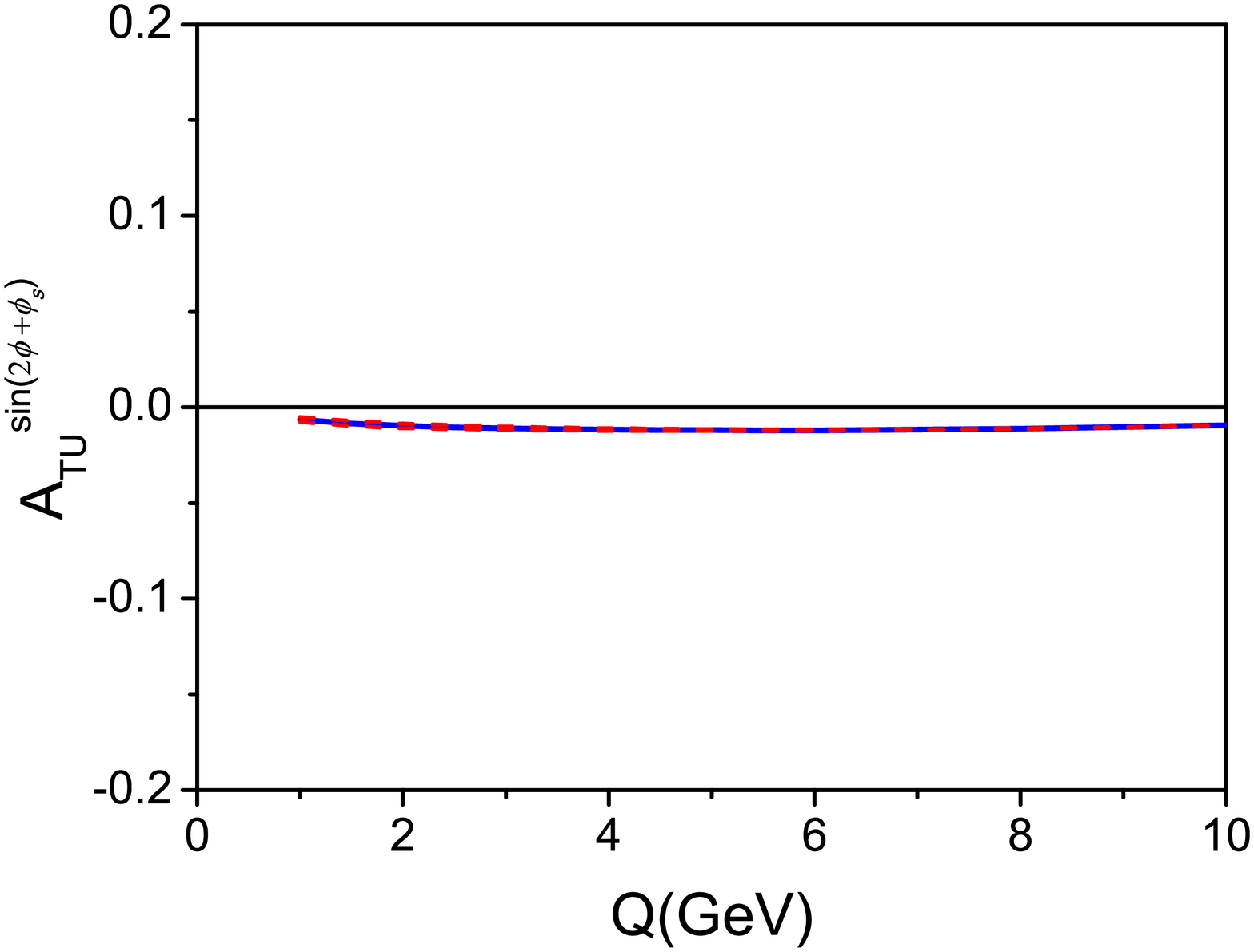}
\caption{The $\sin(2\phi+\phi_S)$ azimuthal asymmetry
$A_{TU}^{\sin(2\phi+\phi_S)}$ depending on $Q$ of target deuteron
polarized $pd$ Drell--Yan process with both $\gamma^*$ and $Z$ taken
into account.}
\label{spdcont2b}       
  \includegraphics[width=0.5\textwidth]{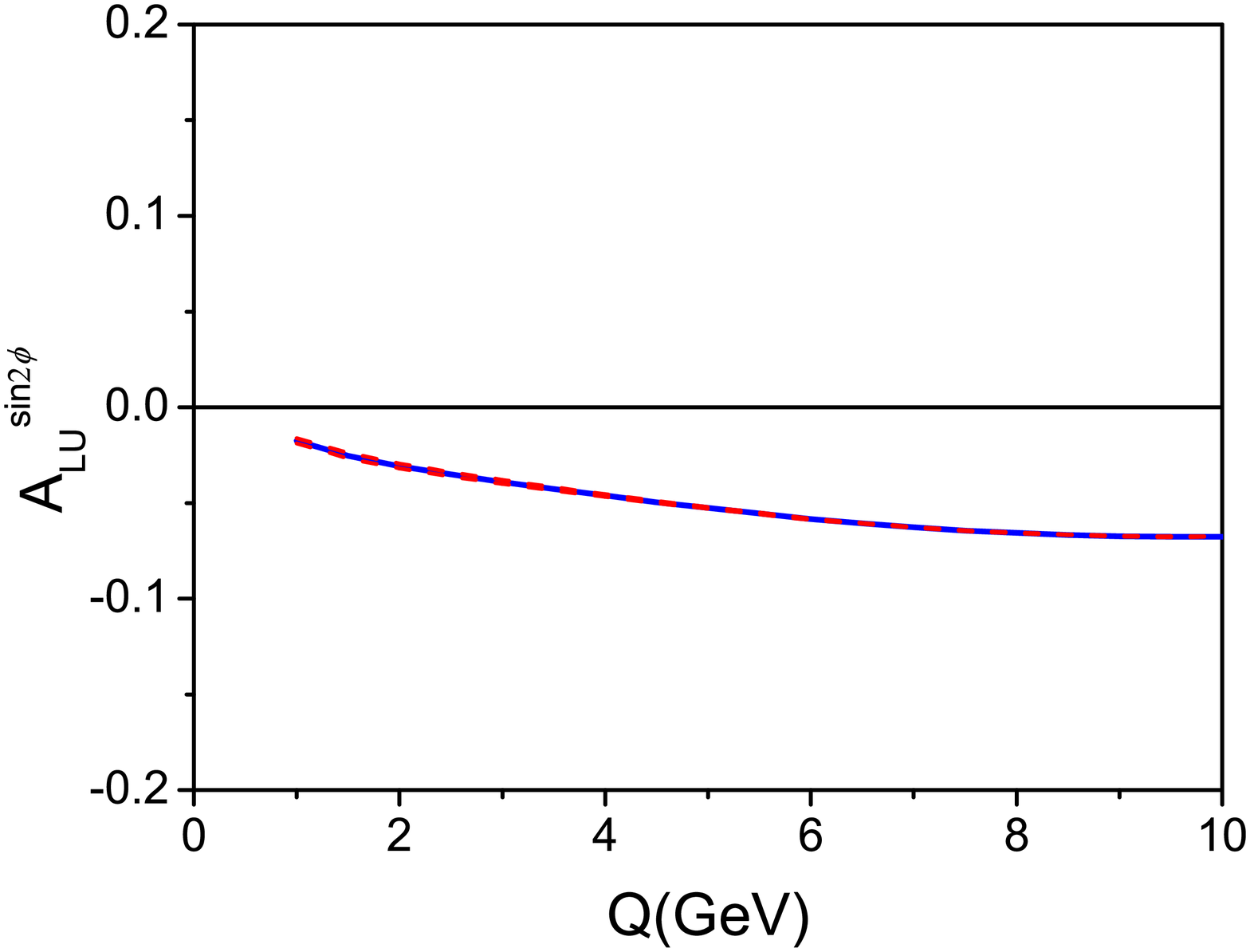}
\caption{The $\sin2\phi$ azimuthal asymmetry $A_{LU}^{\sin2\phi}$ depending on $Q$ of target deuteron
polarized $pd$ Drell--Yan process with both $\gamma^*$ and $Z$ taken into account.}
\label{spdcont2c}       
\end{figure}

\begin{figure}
  \includegraphics[width=0.5\textwidth]{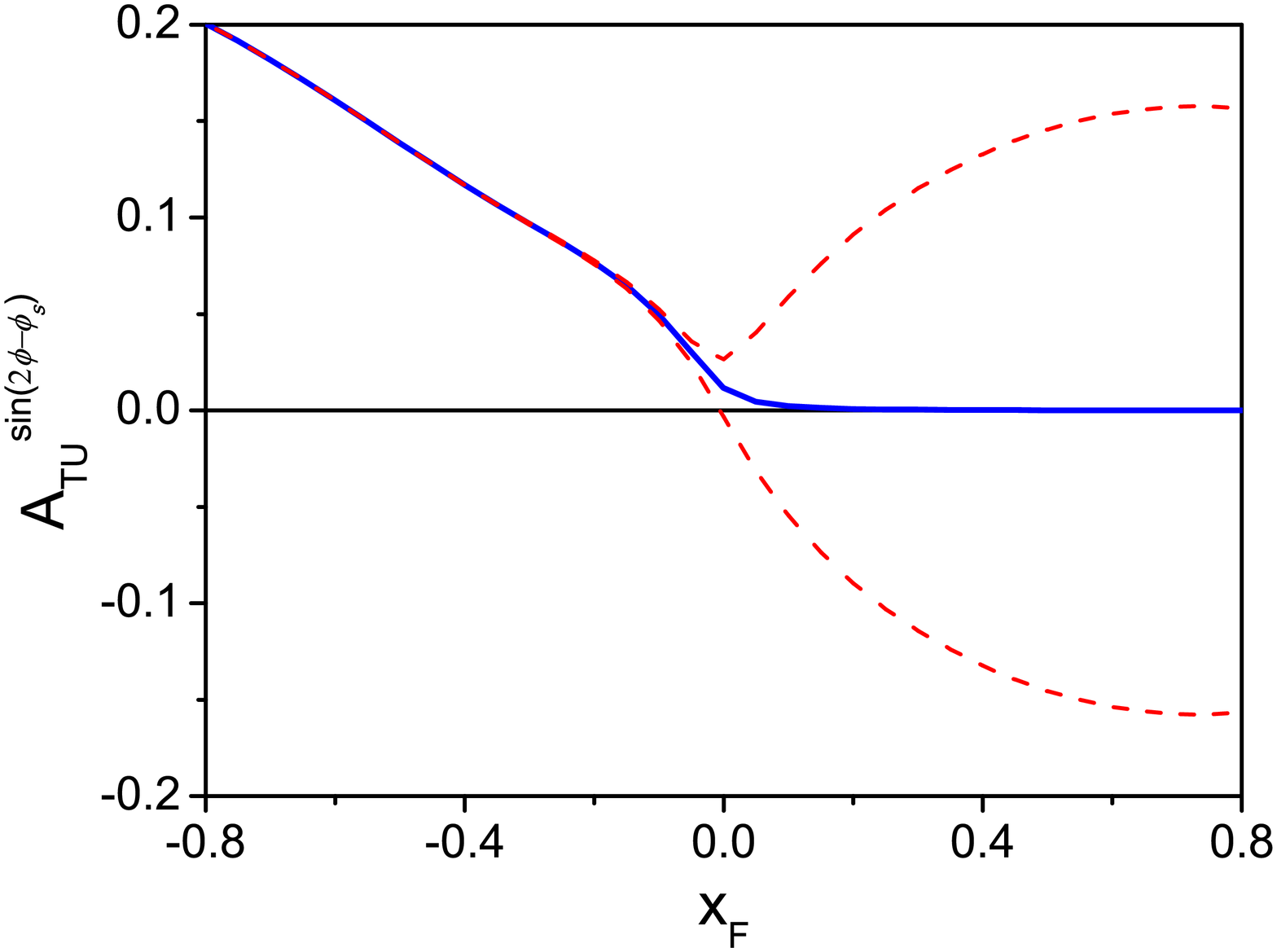}
\caption{The $\sin(2\phi-\phi_S)$ azimuthal asymmetry
$A_{TU}^{\sin(2\phi-\phi_S)}$ depending on $x_F$ of target proton
polarized $pp$ Drell--Yan process at $Q=2$ GeV.}
\label{sppcontxf2a}       
  \includegraphics[width=0.5\textwidth]{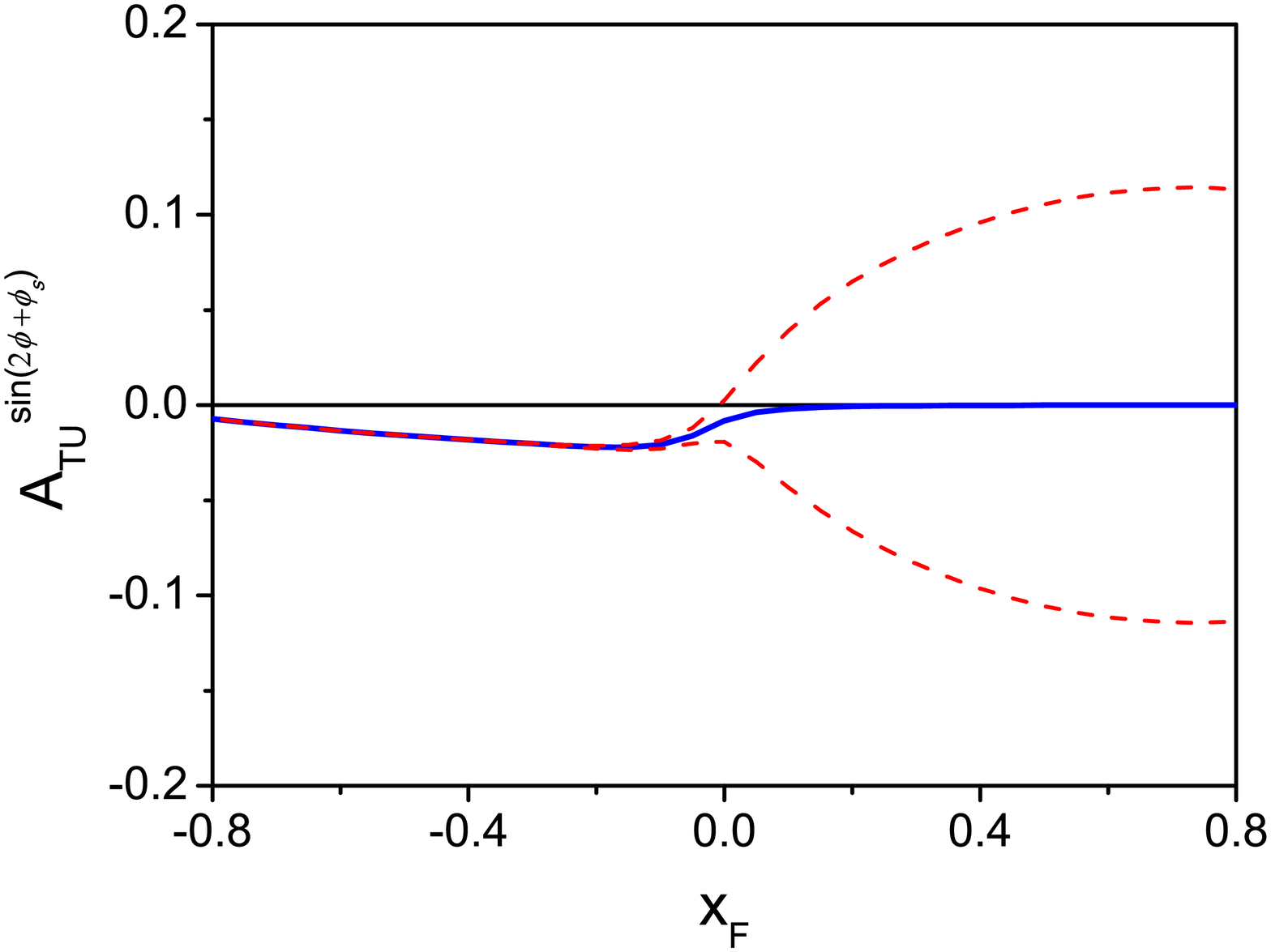}
\caption{The $\sin(2\phi+\phi_S)$ azimuthal asymmetry
$A_{TU}^{\sin(2\phi+\phi_S)}$ depending on $x_F$ of target proton
polarized $pp$ Drell--Yan process at $Q=2$ GeV.}
\label{sppcontxf2b}       
  \includegraphics[width=0.5\textwidth]{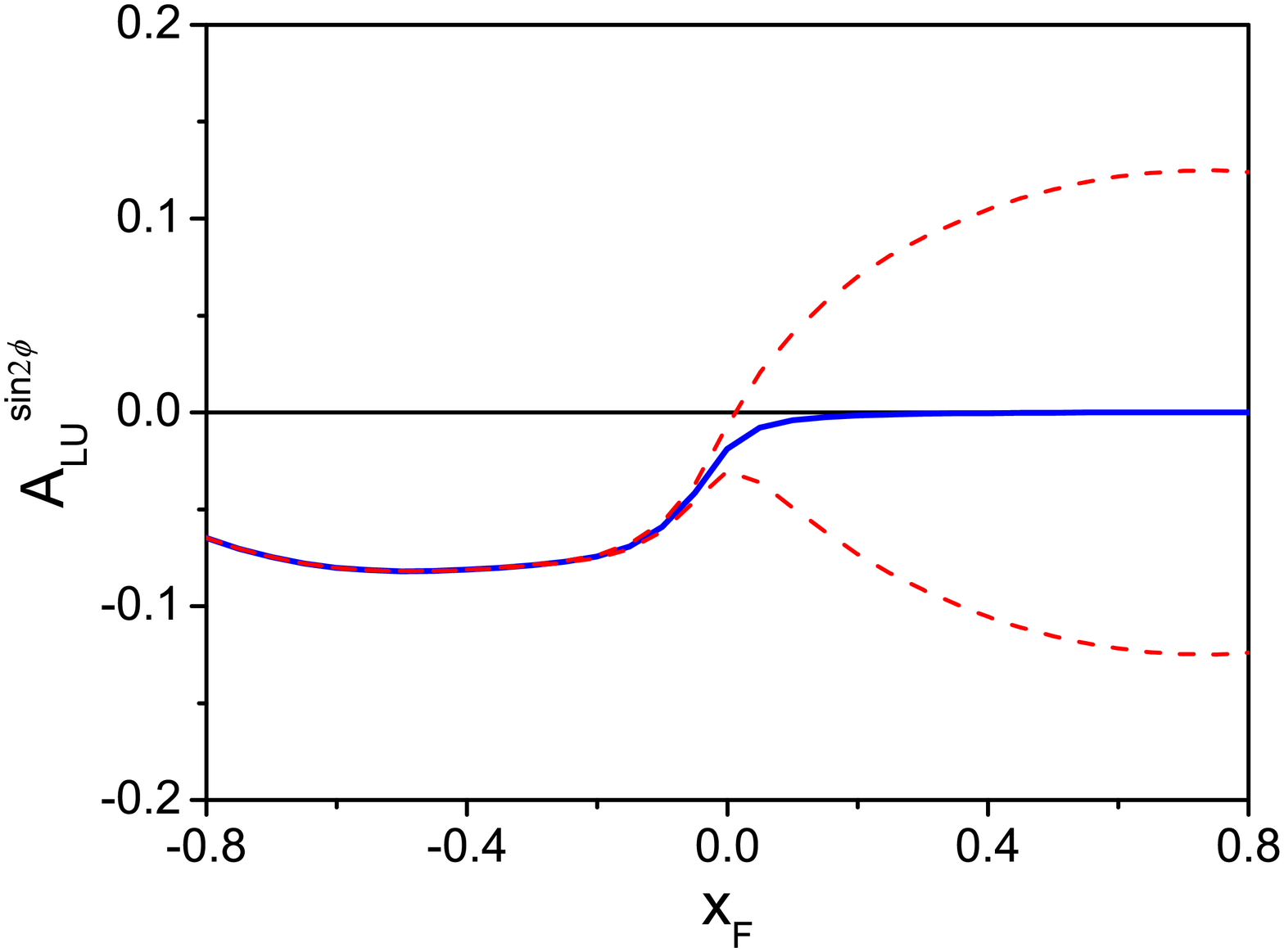}
\caption{The $\sin2\phi$ azimuthal asymmetry $A_{LU}^{\sin2\phi}$ depending on $x_F$ of target proton polarized $pp$ Drell--Yan process at $Q=2$ GeV.}
\label{sppcontxf2c}       
\end{figure}
\begin{figure}
  \includegraphics[width=0.5\textwidth]{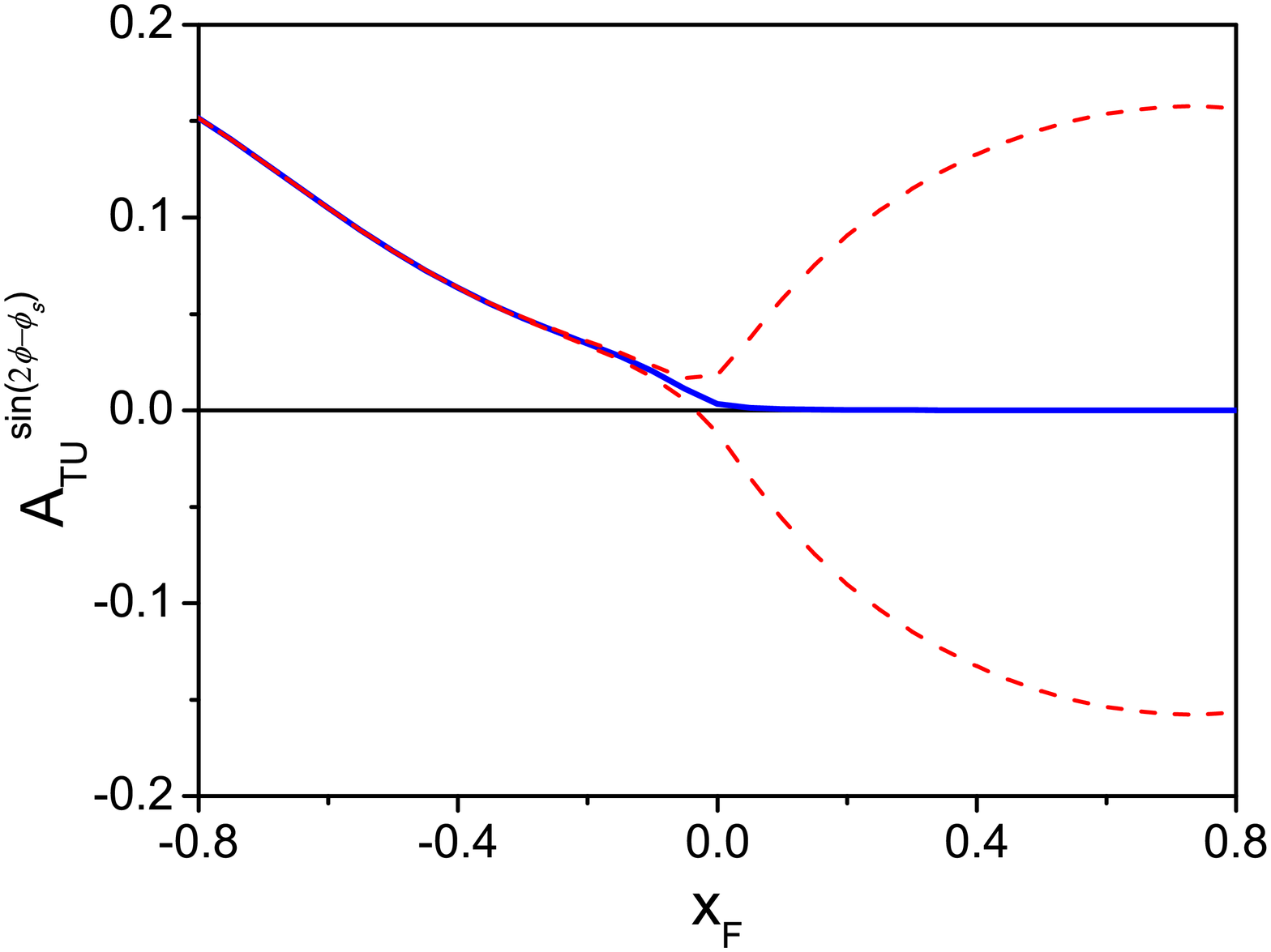}
\caption{The $\sin(2\phi-\phi_S)$ azimuthal asymmetry
$A_{TU}^{\sin(2\phi-\phi_S)}$ depending on $x_F$ of target deuteron
polarized $pd$ Drell--Yan process at $Q=2$ GeV.}
\label{spdcontxf2a}       
  \includegraphics[width=0.5\textwidth]{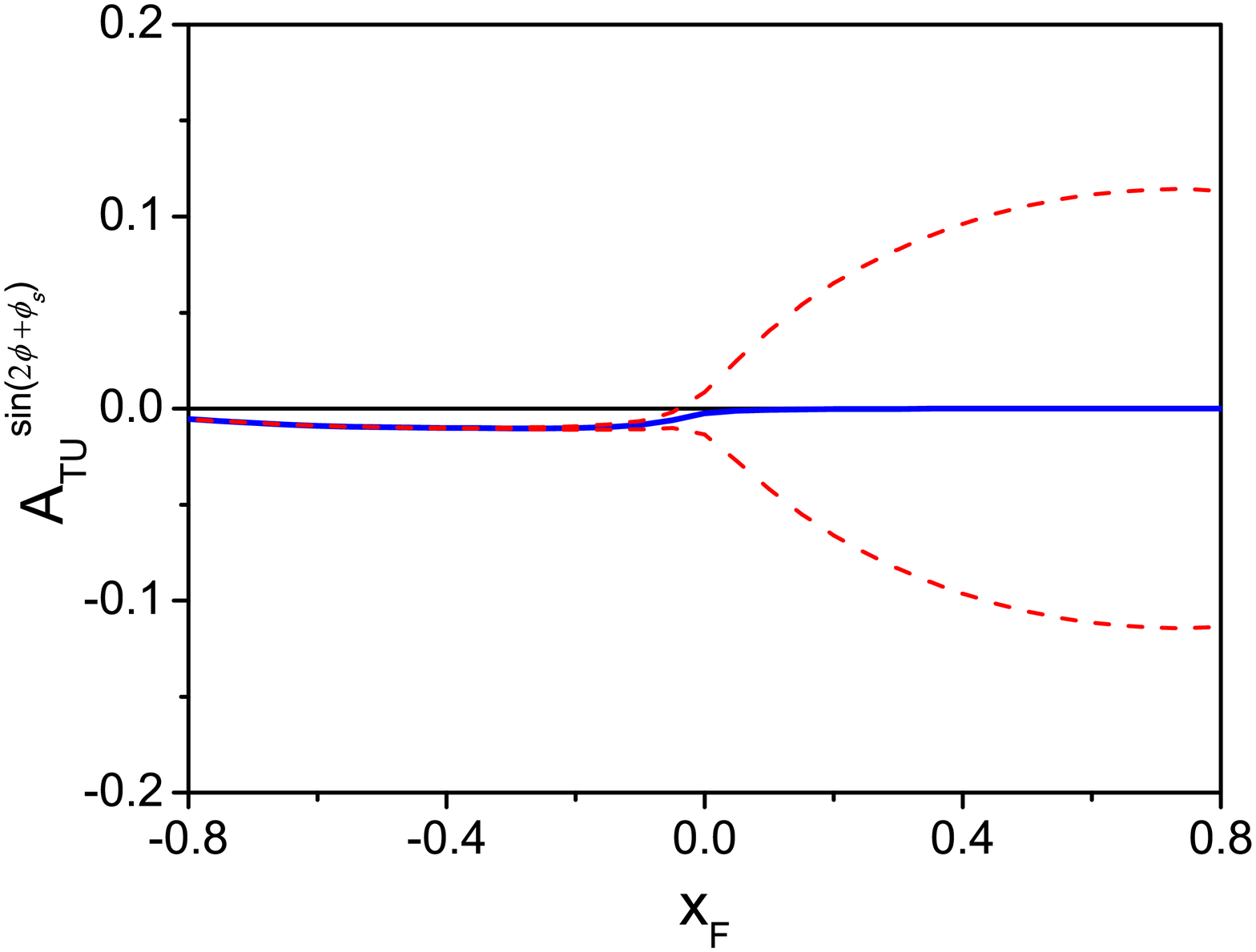}
\caption{The $\sin(2\phi+\phi_S)$ azimuthal asymmetry
$A_{TU}^{\sin(2\phi+\phi_S)}$ depending on $x_F$ of target deuteron
polarized $pd$ Drell--Yan process at $Q=2$ GeV.}
\label{spdcontxf2b}       
  \includegraphics[width=0.5\textwidth]{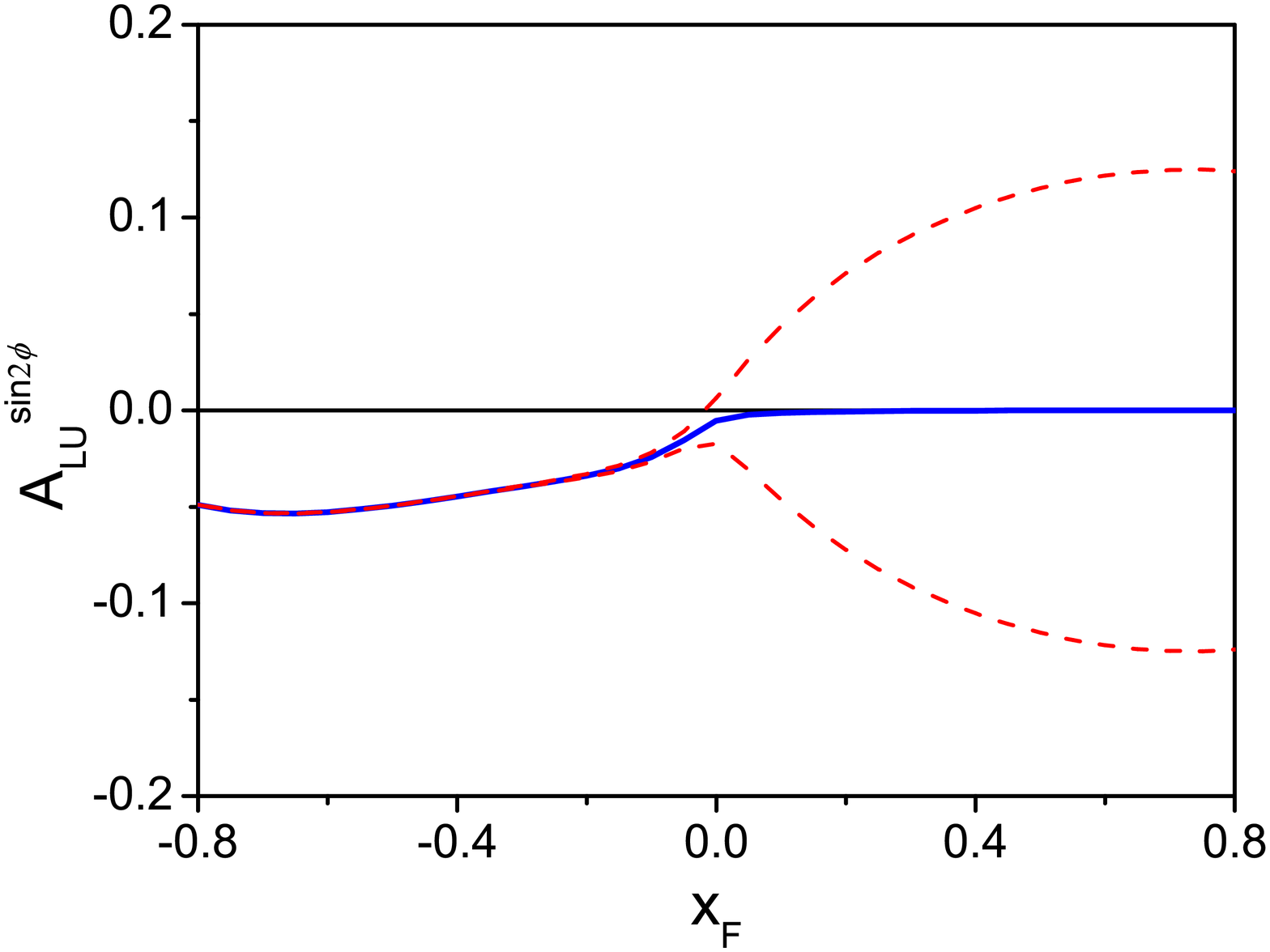}
\caption{The $\sin2\phi$ azimuthal asymmetry $A_{LU}^{\sin2\phi}$ depending on $x_F$ of target deuteron
polarized $pd$ Drell--Yan process at $Q=2$ GeV.}
\label{spdcontxf2c}       
\end{figure}

\begin{figure}
  \includegraphics[width=0.5\textwidth]{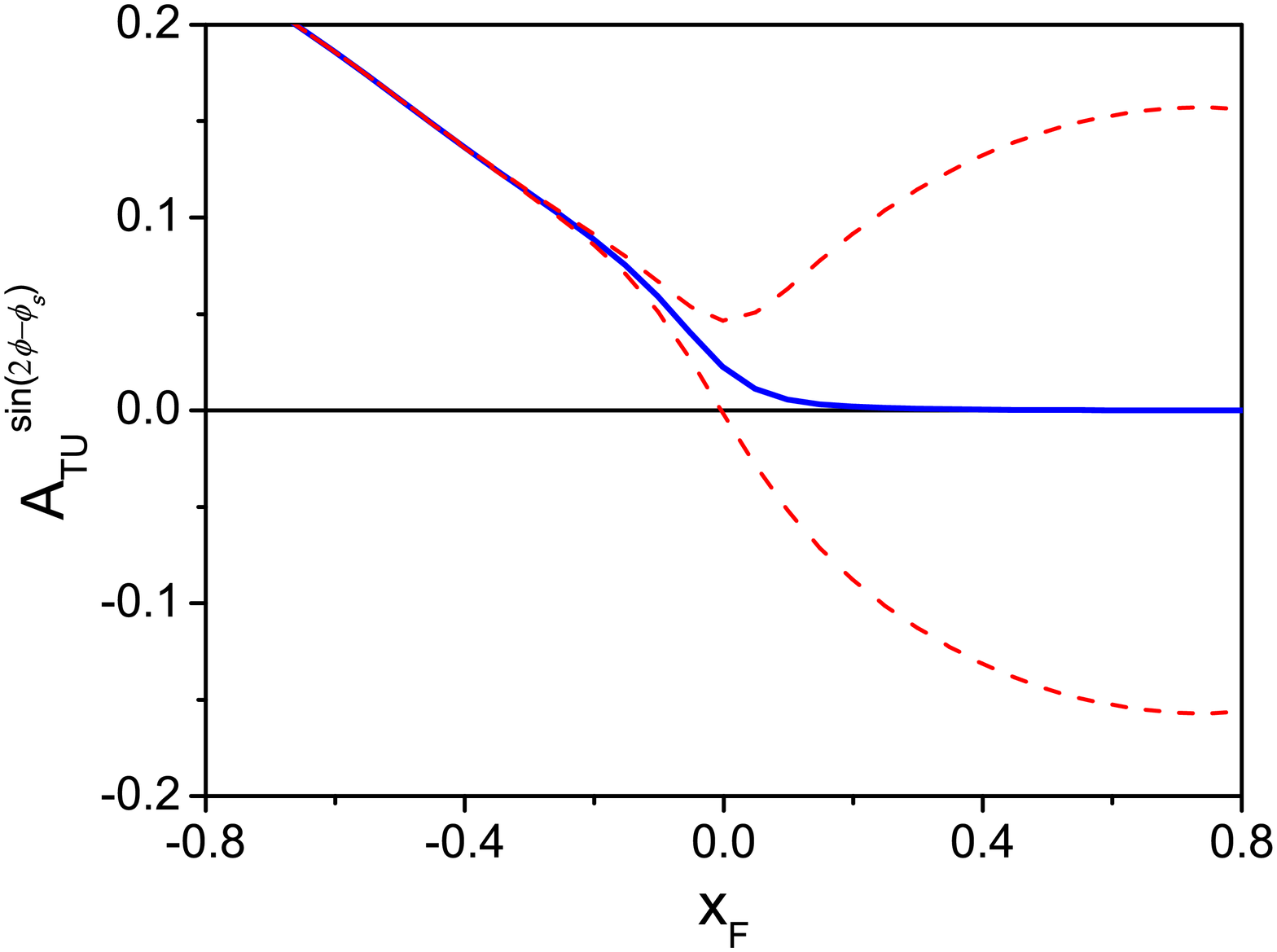}
\caption{The $\sin(2\phi-\phi_S)$ azimuthal asymmetry
$A_{TU}^{\sin(2\phi-\phi_S)}$ depending on $x_F$ of target proton
polarized $pp$ Drell--Yan process at $Q=5$ GeV.}
\label{sppcontxf5a}       
  \includegraphics[width=0.5\textwidth]{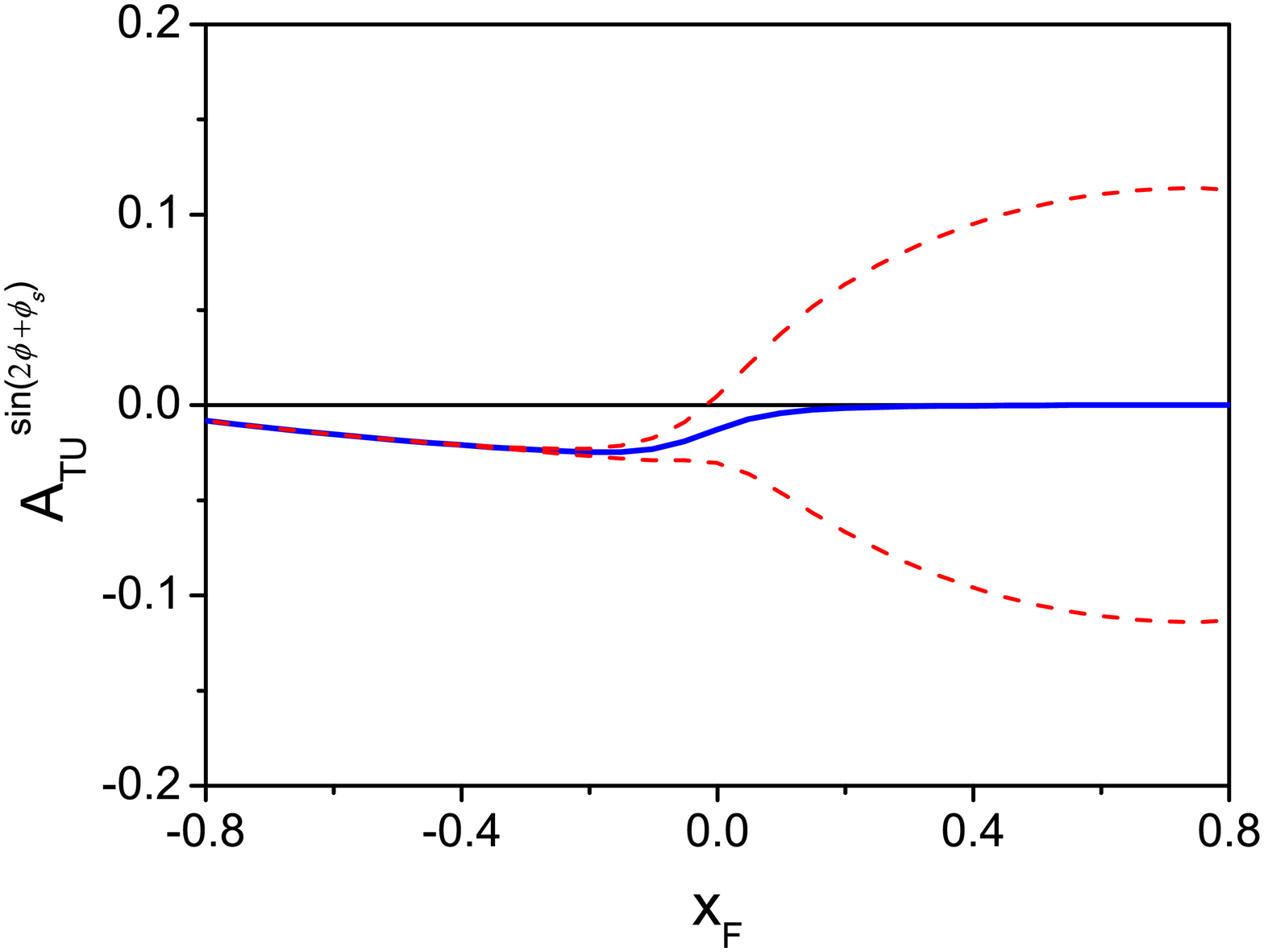}
\caption{The $\sin(2\phi+\phi_S)$ azimuthal asymmetry
$A_{TU}^{\sin(2\phi+\phi_S)}$ depending on $x_F$ of target proton
polarized $pp$ Drell--Yan process at $Q=5$ GeV.}
\label{sppcontxf5b}       
  \includegraphics[width=0.5\textwidth]{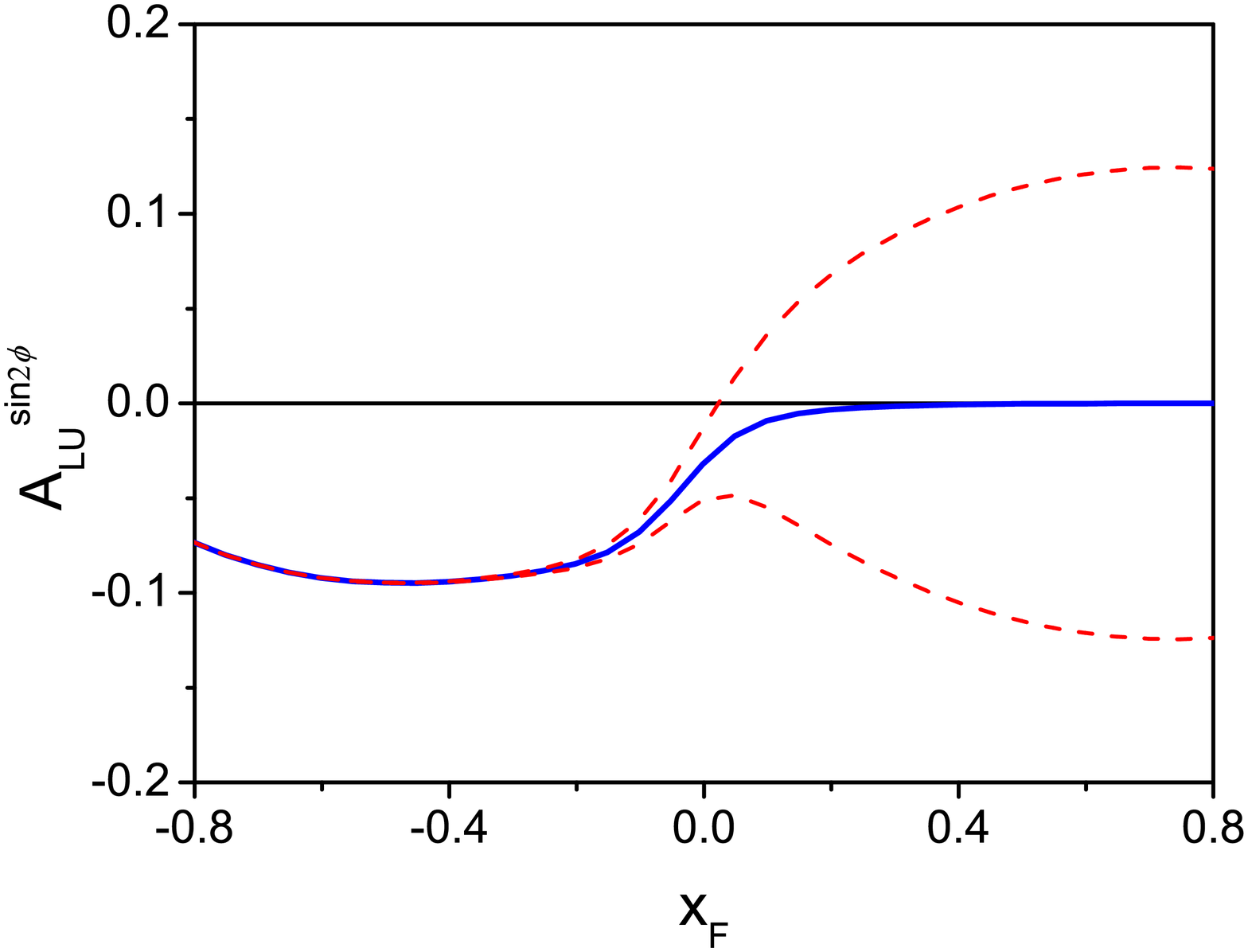}
\caption{The $\sin2\phi$ azimuthal asymmetry $A_{LU}^{\sin2\phi}$ depending on $x_F$ of target proton polarized $pp$ Drell--Yan process at $Q=5$ GeV.}
\label{sppcontxf5c}       
\end{figure}
\begin{figure}
  \includegraphics[width=0.5\textwidth]{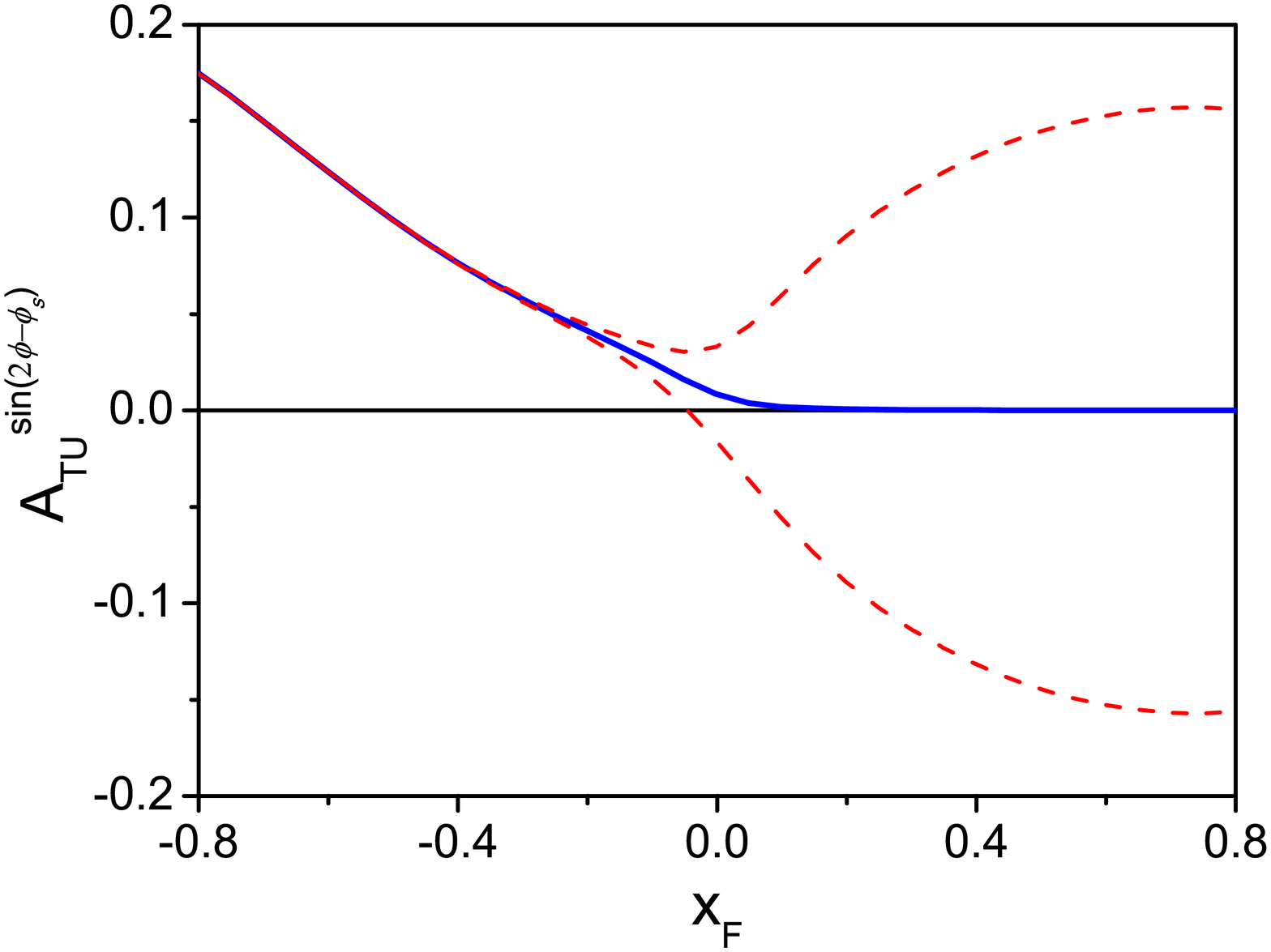}
\caption{The $\sin(2\phi-\phi_S)$ azimuthal asymmetry
$A_{TU}^{\sin(2\phi-\phi_S)}$ depending on $x_F$ of target deuteron
polarized $pd$ Drell--Yan process at $Q=5$ GeV.}
\label{spdcontxf5a}       
  \includegraphics[width=0.5\textwidth]{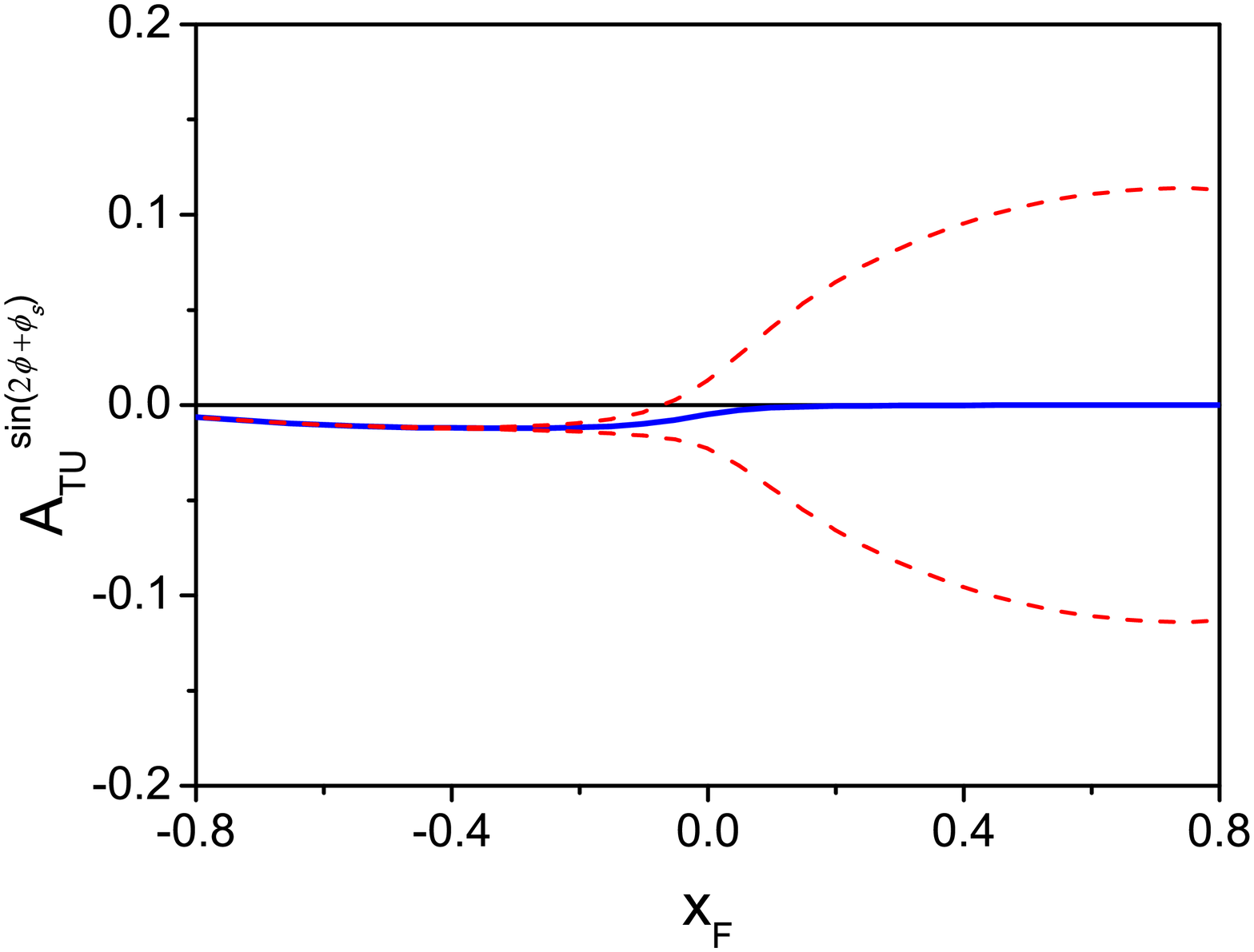}
\caption{The $\sin(2\phi+\phi_S)$ azimuthal asymmetry
$A_{TU}^{\sin(2\phi+\phi_S)}$ depending on $x_F$ of target deuteron
polarized $pd$ Drell--Yan process at $Q=5$ GeV.}
\label{spdcontxf5b}       
  \includegraphics[width=0.5\textwidth]{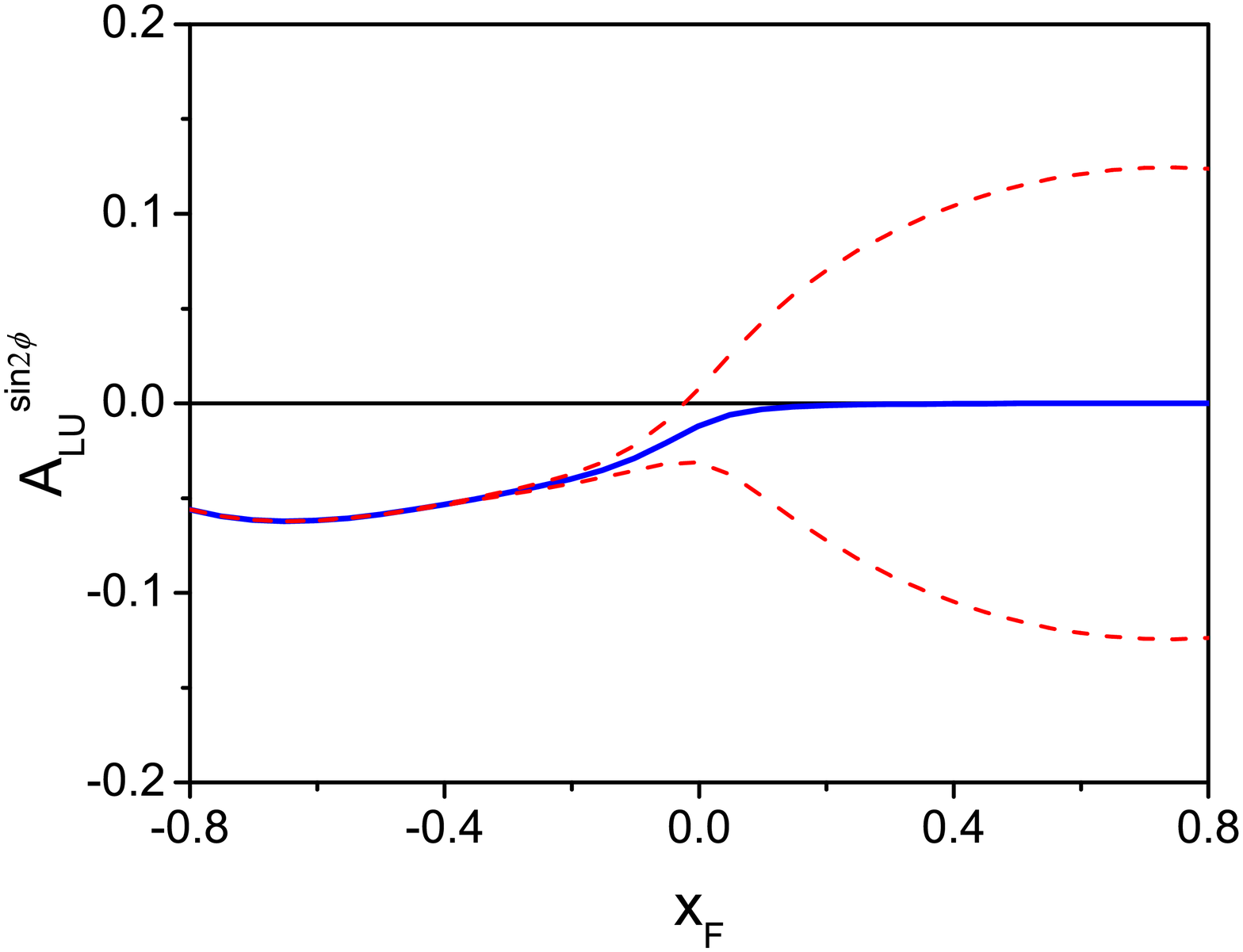}
\caption{The $\sin2\phi$ azimuthal asymmetry $A_{LU}^{\sin2\phi}$ depending on $x_F$ of target deuteron
polarized $pd$ Drell--Yan process at $Q=5$ GeV.}
\label{spdcontxf5c}       
\end{figure}

\begin{figure}
  \includegraphics[width=0.5\textwidth]{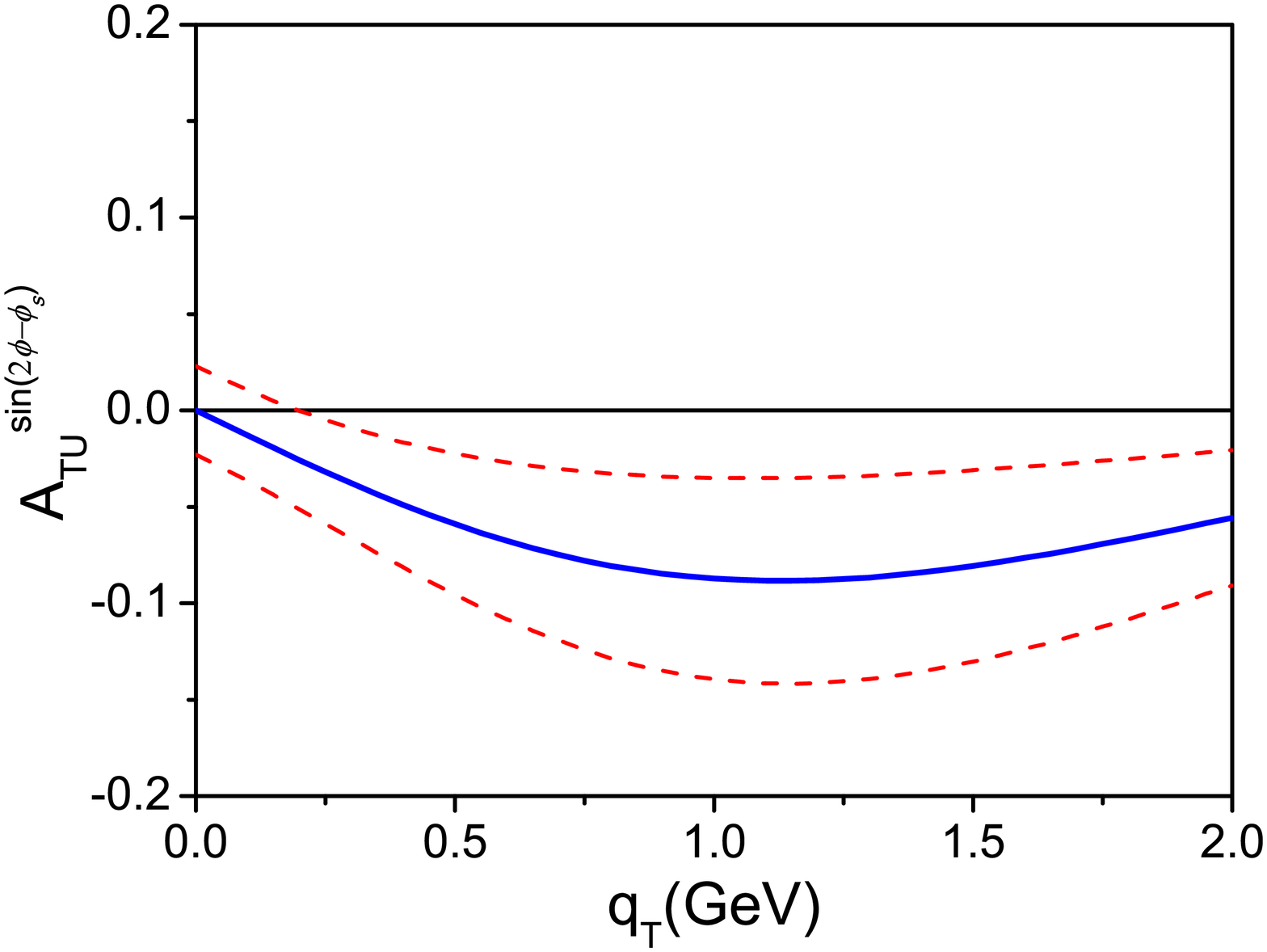}
\caption{The $\sin(2\phi-\phi_S)$ azimuthal asymmetry $A_{TU}^{\sin(2\phi-\phi_S)}$ depending on $q_T$ of target proton polarized $pp$ dilepton production process at the $Z$ pole.}
\label{sppza}       
  \includegraphics[width=0.5\textwidth]{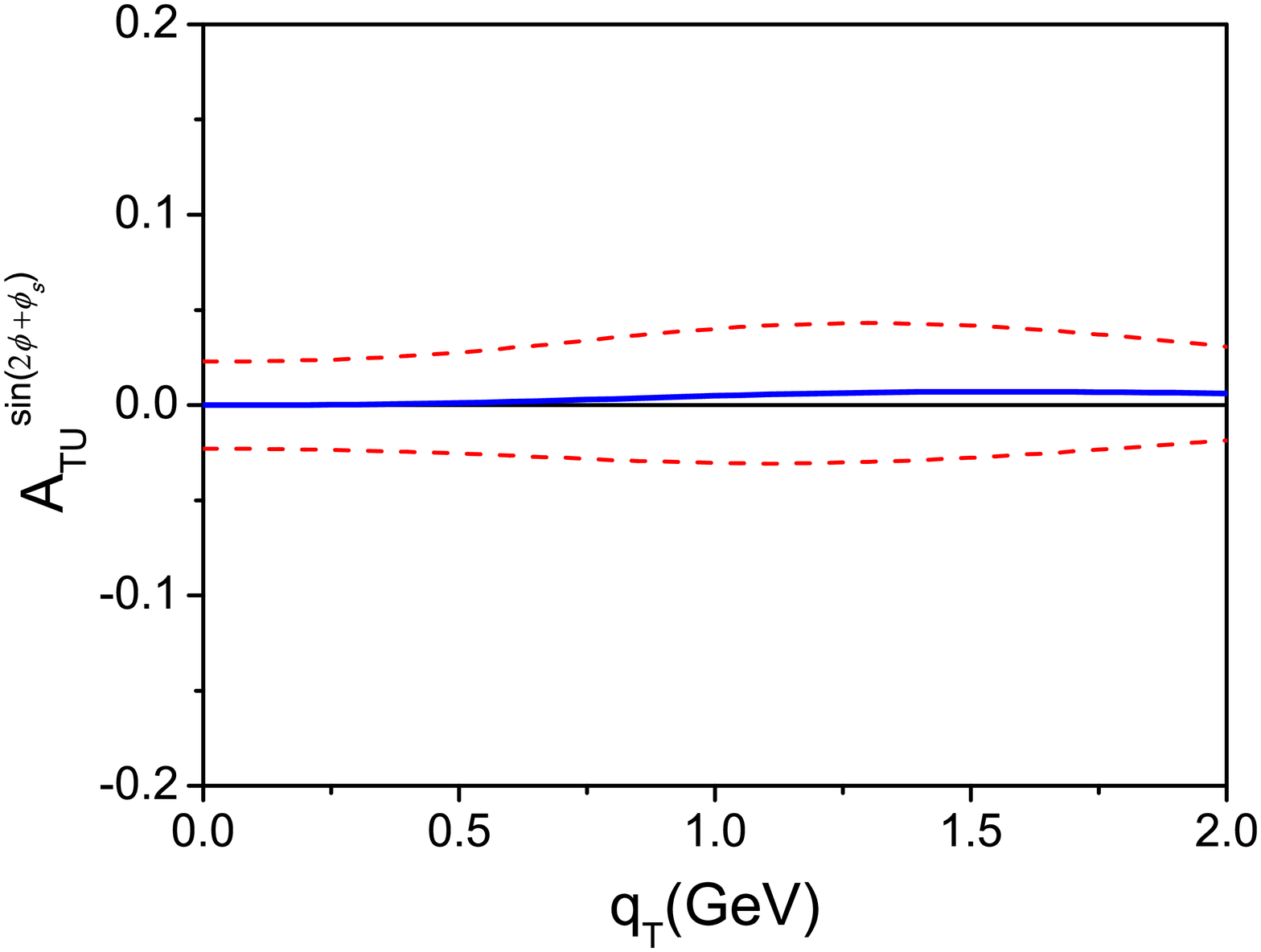}
\caption{The $\sin(2\phi+\phi_S)$ azimuthal asymmetry $A_{TU}^{\sin(2\phi+\phi_S)}$ depending on $q_T$ of target proton polarized $pp$  dilepton production process at the $Z$ pole.}
\label{sppzb}       
  \includegraphics[width=0.5\textwidth]{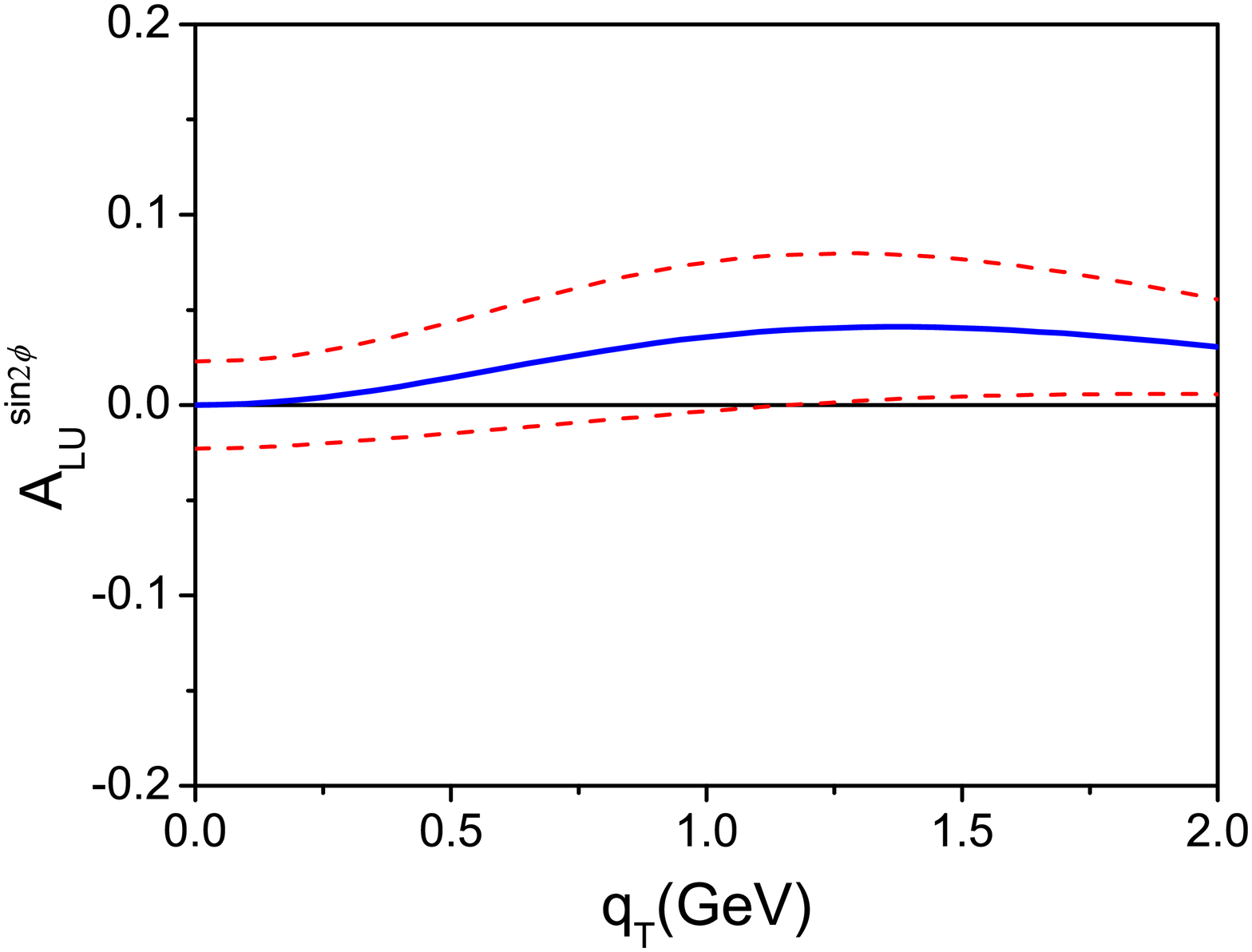}
\caption{The $\sin2\phi$ azimuthal asymmetry $A_{TU}^{\sin2\phi}$ depending on $q_T$ of target proton polarized $pp$  dilepton production process at the $Z$ pole.}
\label{sppzc}       
\end{figure}
\begin{figure}
  \includegraphics[width=0.5\textwidth]{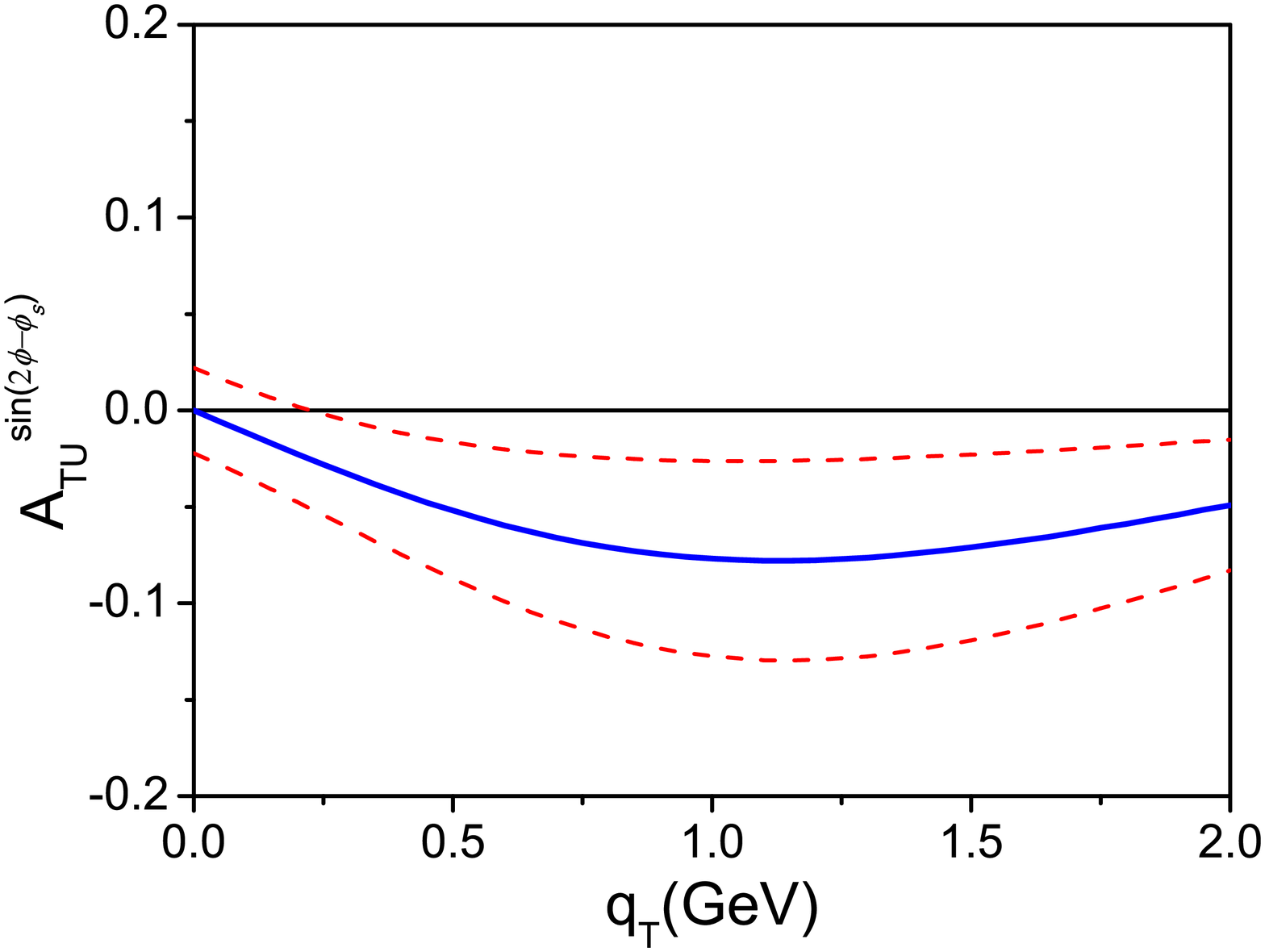}
\caption{The $\sin(2\phi-\phi_S)$ azimuthal asymmetry
$A_{TU}^{\sin(2\phi-\phi_S)}$ depending on $q_T$ of target deuteron
polarized $pd$ dilepton production process at the $Z$ pole.}
\label{spdza}       
  \includegraphics[width=0.5\textwidth]{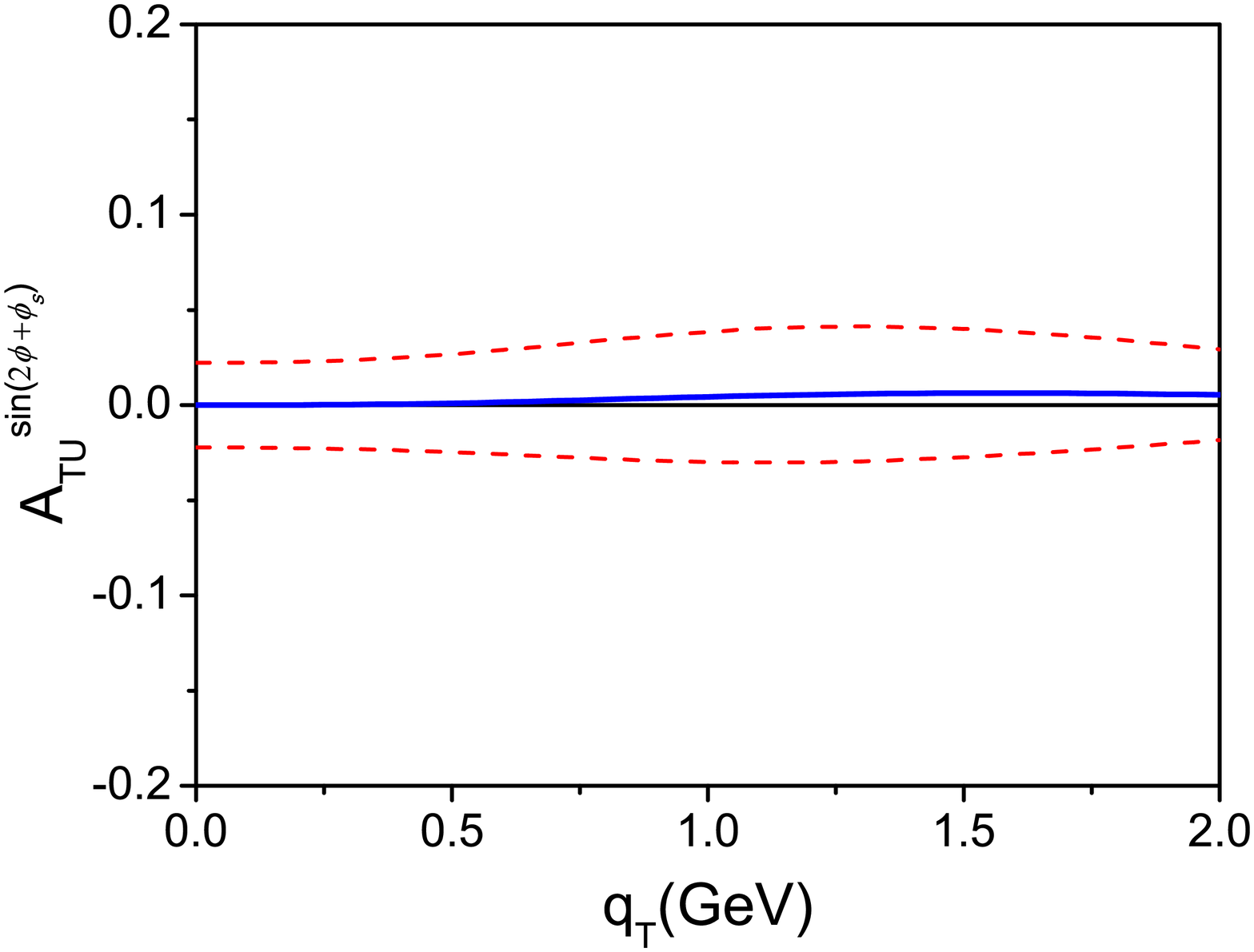}
\caption{The $\sin(2\phi+\phi_S)$ azimuthal asymmetry
$A_{TU}^{\sin(2\phi+\phi_S)}$ depending on $q_T$ of target deuteron
polarized $pp$  dilepton production process at the $Z$ pole.}
\label{spdzb}       
  \includegraphics[width=0.5\textwidth]{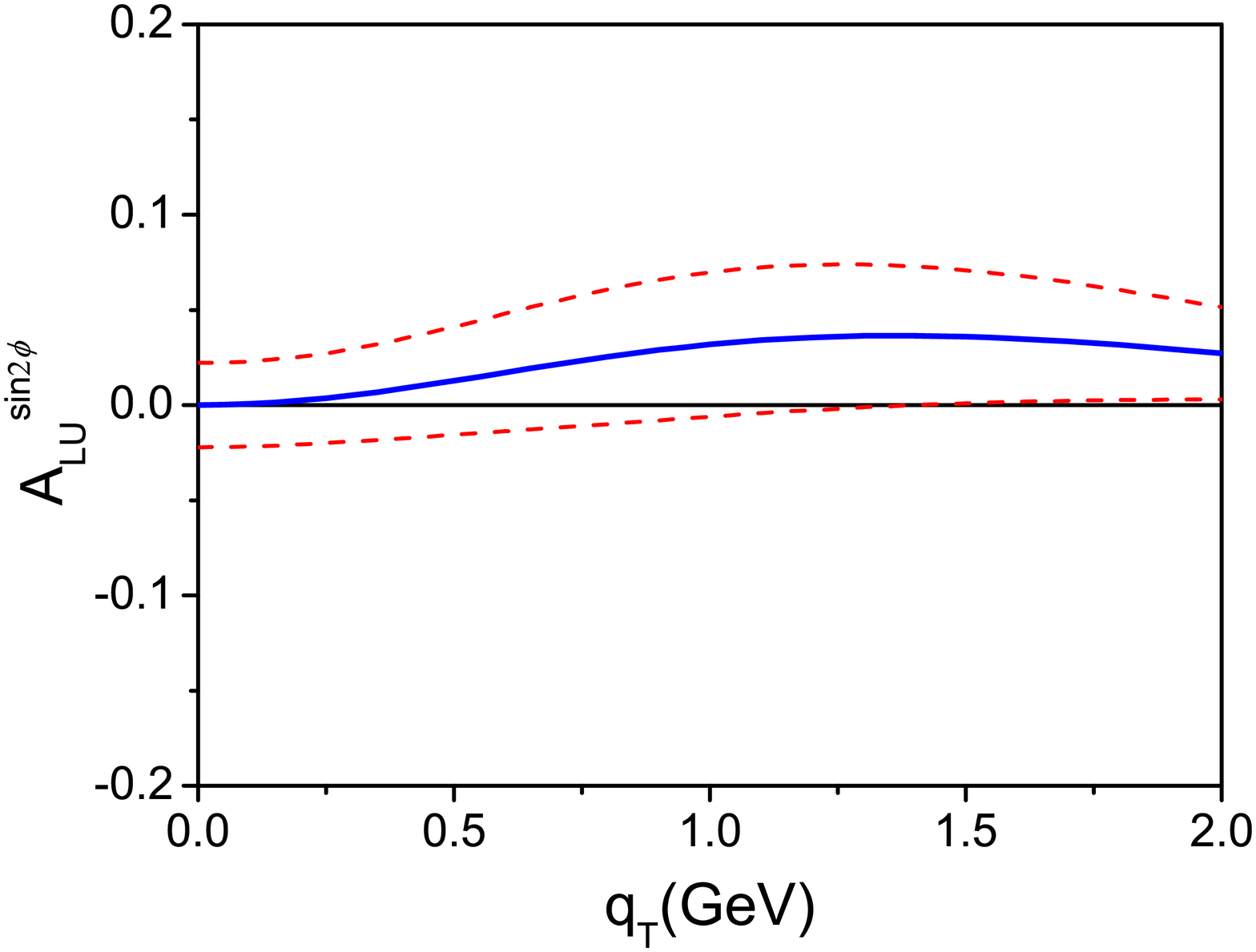}
\caption{The $\sin2\phi$ azimuthal asymmetry $A_{TU}^{\sin2\phi}$
depending on $q_T$ of target deuteron polarized $pd$  dilepton
production process at the $Z$ pole.}
\label{spdzc}       
\end{figure}

To calculate the SSAs of the $pp$ and $pd$ dilepton production processes at AFTER,
we also need the distribution functions $h_{1T}^\perp$, $h_{1L}^\perp$ and $h_1$ besides the Boer--Mulders function $h_1^\perp$. In our calculation, we adopt ansatz of these $T$-even distribution functions calculated from the light-cone quark-diquark model. In this model, the Melosh--Wigner rotation,
which is important to understand the proton spin puzzle due to the relativistic effect of quark transversal motions~\cite{Wigner:1939cj,Melosh:1974cu,MaOld1,MaOld2}, is taken into account. This model has been applied to calculate helicity distributions~\cite{Ma:1996sr},
transversity distributions~\cite{Ma:1997gy,Schmidt:1997vm} and some other 3dPDFs or TMDs~\cite{She:2009jq,Zhu:2011zza,Lu:2004au}, and has been used to analysis related azimuthal
spin asymmetries in SIDIS processes~\cite{Ma:2001ie,Ma:2002ns}. The model results of these distribution functions are expressed as~\cite{She:2009jq,Zhu:2011zza,Ma:1997gy,Schmidt:1997vm}
\begin{align}
j_u^{v}(x,\bm{k}_T^2) =&\big[f_{1u}^{v}(x,\bm{k}_T^2)-\frac{1}{2}f_{1d}^{v}(x,\bm{k}_T^2)\big]W_S^j(x,\bm{k}_T^2)\nonumber\\
&-\frac{1}{6}f_{1d}^{v}(x,\bm{k}_T^2)W_V^j(x,\bm{k}_T^2),\\
j_d^{v}(x,\bm{k}_T^2) =&-\frac{1}{3}f_{1d}^{v}(x,\bm{k}_T^2)W_V^j(x,\bm{k}_T^2),
\label{tevenpdf}
\end{align}
where $j=h_1$, $h_{1T}^\perp$, $h_{1L}^\perp$, and the superscript $v$ stands for valence quark distributions.
The factors $W_{S/V}^j(x,\bm{k}_T^2)$ are the Melosh--Wigner rotations for scalar or axial vector spectator-diquark respectively,
having the forms:
\begin{align}
W_D^{h_1}(x,\bm{k}_T^2)&= \frac{(x\mathcal{M}_D + m_q)^2}{(x\mathcal{M}_D + m_q)^2 +\bm{k}_T^2},\\
W_D^{h_{1T}^\perp}(x,\bm{k}_T^2)&= -\frac{2m_N^2}{(x\mathcal{M}_D + m_q)^2 + \bm{k}_T^2},\\
W_D^{h_{1L}^\perp}(x,\bm{k}_T^2)&= -\frac{2m_N(x\mathcal{M}_D + m_q)}{(x\mathcal{M}_D + m_q)^2 +\bm{k}_T^2},
\end{align}
where
\begin{equation}
\mathcal{M}_D = \sqrt{\frac{m_q^2 + \bm{k}_T^2}{x} + \frac{m_D^2 + \bm{k}_T^2}{1- x}}.
\end{equation}

The distribution functions $h_1$, $h_{1T}^\perp$, $h_{1L}^\perp$ of sea quarks are constrained by the positivity bounds~\cite{Bacchetta:1999kz}:
\begin{eqnarray}
\left| \bar{h}_{1q} (x,\bm{k}_T^2) \right| \leq \bar{f}_{1q}(x,\bm{k}_T^2),\label{h1bound}\\
\left| \frac{\bm{k}_T^2}{2m_N^2}\bar{h}_{1Tq}^{\perp}(x,\,k_T^2) \right| \leq \bar{f}_{1q}(x,\bm{k}_T^2), \label{pretzbound}\\
\left| \frac{\bm{k}_T}{m_N}\bar{h}_{1Lq}^{\perp} (x,\bm{k}_T^2)\right| \leq \bar{f}_{1q}(x,\bm{k}_T^2).\label{hbound}
\end{eqnarray}
When considering the effects of these distribution functions of sea quarks contributing to the asymmetries,
we can get the upper and lower limits of the azimuthal asymmetries by saturating the positivity bounds.

In Figs.~\ref{sppcont1a}-\ref{sppcont1c}, we respectively show the
$\sin(2\phi-\phi_S)$, $\sin(2\phi+\phi_S)$ and $\sin2\phi$ azimuthal
asymmetries depending on $Q$ from $2$ GeV to $30$ GeV of the target proton polarized $pp$ Drell--Yan process
at AFTER including $Z$ taken into account,
and the corresponding results for deuteron target polarized $pd$ process are shown in Figs.~\ref{spdcont1a}-\ref{spdcont1c}.
These asymmetries in the far backward region, with rapidity cut $[-4.8,-2]$, are shown in Figs.~\ref{sppcont2a}-\ref{spdcont2c}, with $Q$ running from $1$ GeV to $10$ GeV.
In Figs.~\ref{sppcontxf2a}-\ref{spdcontxf5c}, we respectively show the
$\sin(2\phi-\phi_S)$, $\sin(2\phi+\phi_S)$ and $\sin2\phi$ azimuthal
asymmetries depending on $x_F$ of the target proton and deuteron polarized $pp$ and $pd$ Drell--Yan processes at AFTER
with $Q=2$ GeV and $Q=5$ GeV as for low and mid $Q$ regions.
Figs.~\ref{sppza}--\ref{sppzc} show the
$\sin(2\phi-\phi_S)$, $\sin(2\phi+\phi_S)$ and $\sin2\phi$ azimuthal
asymmetries of the target proton polarized $pp$ process around the
$Z$-pole at AFTER. The corresponding azimuthal asymmetries of $pd$
processes with target deuteron transversally or longitudinally
polarized are shown in Figs.~\ref{spdza}--\ref{spdzc}.

\section{Discussion and conclusions}

In this paper, we calculate the $\cos2\phi$ azimuthal asymmetries of unpolarized $pp$ and $pd$ dilepton production processes in the Drell--Yan continuum region and around the $Z$ resonance region.
We also calculate the $\sin(2\phi-\phi_S)$, $\sin(2\phi+\phi_S)$ and $\sin2\phi$ azimuthal asymmetries of single transversally or longitudinally polarized $pp$ and $pd$ dilepton production processes in these regions.

Our calculations are concentrated on some issues related to the spin
physics part of the AFTER project, a multi-purpose fixed-target
experiment using the proton and lead-ion beams of the LHC extracted
by a bent crystal, proposed by Brodsky, Fleuret, Hadjidakis and
Lansberg~\cite{Brodsky:2012vg}. We present an estimation of the
azimuthal asymmetries for a fixed-target experiment using the LHC
7~TeV proton beams with the proton or deuteron target unpolarized
and transversally or longitudinally polarized. As the target
is conveniently polarized, it is an ideal ground to study the spin
physics at AFTER with $\sqrt{s}=115$ GeV and high luminosity. It is
feasible to measure these azimuthal asymmetries at AFTER.
This will help us to study the three dimensional or transverse momentum
dependent parton distributions (3dPDFs or TMDs), and consequently help understand
and test the QCD and hadron structure at such a high laboratory energy.

\begin{acknowledgements}
This work is partially supported by National Natural
Science Foundation of China (Grants No.~11021092, No.~10975003,
No.~11035003, and No.~11120101004), by the Research Fund for the
Doctoral Program of Higher Education (China)
\end{acknowledgements}



\end{document}